\documentclass[10pt,twocolumn,journal]{IEEEtran}
\usepackage{amsmath,amssymb,amsfonts}
\usepackage{graphicx}
\usepackage{cite}
\usepackage{bm}
\usepackage{color}
\usepackage{algorithm,algpseudocode}
\usepackage{float}

\def\ebf{{\bf e}}

\def\pbf{{\bf p}}

\def\rbf{{\bf r}}

\def\vbf{{\bf v}}
\def\wbf{{\bf w}}
\def\xbf{{\bf x}}

\def\zbf{{\bf z}}
\def\rbf{{\bf r}}
\def\xbf{{\bf x}}

\def\Gbf{{\bf G}}

\def\Vbf{{\bf V}}

\def\Zbf{{\bf Z}}

\newtheorem{proposition}{Proposition}

\begin{document}

\IEEEoverridecommandlockouts

\title{Multi-user Multi-task Offloading and Resource Allocation in Mobile Cloud Systems
}

\author{
Meng-Hsi Chen, Ben Liang, \IEEEmembership{Fellow, IEEE}, Min Dong, \IEEEmembership{Senior Member, IEEE}
\thanks{Meng-Hsi Chen and
Ben Liang are with the Department of Electrical and Computer
Engineering, University of Toronto, Toronto, Canada (e-mail:
$\{$mchen, liang$\}$@ece.utoronto.ca).

Min Dong is with the Department of Electrical, Computer and Software
Engineering, University of Ontario Institute of Technology, Oshawa,
Canada (e-mail: min.dong@uoit.ca).}

\thanks{This work has been funded in part by a Natural Sciences and Engineering Research
Council (NSERC) of Canada Strategic Project Grant and in part by
NSERC Discovery Grants. }
\thanks{
 Preliminary result of this work has
appeared in \cite{chen2016icc} and \cite{chen2017infocom}. }
 }

\maketitle \vspace{-1cm}

\begin{abstract}
We consider a general multi-user Mobile Cloud Computing (MCC) system where each mobile user has multiple independent tasks. These mobile users share the computation and communication resources while offloading tasks to the cloud. We study both the conventional MCC where tasks are offloaded to the cloud through a wireless access point, and MCC with a computing access point (CAP), where the CAP serves both as the network access gateway and a computation service provider to the mobile users. We aim to jointly optimize the offloading decisions of all users as well as the allocation of computation and communication resources, to minimize the overall cost of energy, computation, and delay for all users. The optimization problem is formulated as a non-convex quadratically constrained quadratic program, which is NP-hard in general. For the case without a CAP, an efficient approximate solution named MUMTO is proposed by using separable semidefinite relaxation (SDR), followed by recovery of the binary offloading decision and optimal allocation of the communication resource. To solve the more complicated problem with a CAP, we further propose an efficient three-step algorithm named MUMTO-C comprising of generalized MUMTO SDR with CAP, alternating optimization, and sequential tuning, which always computes a locally optimal solution. For performance benchmarking, we further present numerical lower bounds of the minimum system cost with and without the CAP. By comparison with this lower bound, our simulation results show that the proposed solutions for both scenarios give nearly optimal performance under various parameter settings, and the resultant efficient utilization of a CAP can bring substantial cost benefit.
\end{abstract}



\section{Introduction} \label{}
Mobile Cloud Computing (MCC) brings abundant cloud
resources to extend the capabilities of resource-limited mobiles
devices to improve the user experience
\cite{kumar2013}\cite{dinh2013}. With the help of cloud resources,
mobile devices can potentially reduce their energy consumption or
processing delay by offloading their tasks to the cloud. However,
integration between mobile devices and the cloud may affect the
quality of service of those offloaded tasks and overall mobile
device energy usage due to additional communication and computation
delays and transceiver energy consumption.

The case of a single mobile user offloading its entire application
to the cloud was studied in \cite{kumar2010,wen2012}. Furthermore,
offloading by multiple mobile users was considered in
\cite{chen2014,ren2013,meskar2017}, where each user has a single
application or task to be offloaded to the cloud in its entirety.
Different from such whole-application offloading, the authors of
\cite{cuervo2010,
chun2011,kosta2012,zhang2013infocom,kao2015,mahmoodi2016,wu2016}
considered partitioning an application into multiple tasks.
 In all these cases, the partitioning problem
results in integer programming that is NP-hard in general.

In conventional MCC systems, communication between the mobile
devices and the remote cloud server often is over a long distance,
which may result in large communication delay in task offloading. In
contrast, with an aim to reduce the communication delay for those
offloaded tasks, Mobile/Multiaccess Edge Computing (MEC), as defined
by the European Telecommunications Standards Institute, refers to a
distributed MCC system where computing resources are installed
locally at or near the base station of a cellular
network\cite{etsi2016framework,liang2017,mao2017}. MEC shares
similarities with micro cloud centers \cite{greenberg2008},
cloudlets\cite{satyanarayanan2009}, fog computing \cite{bonomi2012},
and cyber-foraging\cite{lewis2015}, except that the MEC computing
servers are managed by a mobile service provider, which allows more
direct control and resource management. Similar to the concept of
MEC, one may define a general computing access point (CAP), which is
a wireless access point or a cellular base station with built-in
computation capability to serve the mobile users' computing tasks. These tasks may be
processed locally at the mobile devices, sent to the CAP, or further
forwarded to a remote cloud server. With the additional option of
computation by the CAP, we can reduce the need for access to the
remote cloud server, hence decreasing the communication delay and
also potentially the overall energy and computation cost.


In this work, we study the joint optimization of task offloading and
resource allocation in a general mobile cloud access network
consisting of multiple mobile users, each having multiple
independent tasks. The wireless access point may serve its
conventional networking function and only forward the received tasks
to the remote cloud server, or it may be a CAP that additionally has
limited built-in computation capability to directly process some of
the tasks by itself. We take into consideration the computation and
communication energies, CAP and cloud usage costs, and communication
and processing delays at local user devices, the CAP, and the remote
cloud server.

 The multi-user multi-task scenario adds substantial
challenge to system design, since we need to jointly consider both
the offloading decisions and the sharing of limited computation and
communication resources among all users as they compete with each
other while offloading tasks. In particular, the delays of the
offloaded tasks of a user will be affected by its assigned
computation and communication resources, as well as the scheduling of those tasks in the computation and communication pipelines.
 Therefore, scheduling the tasks of even a single user contains a multi-machine flow-shop problem \cite{Garey1976}, which has no known optimal
 solution in the literature. In this work, we propose efficient heuristic solutions based on semi-definite relaxation methods,
 together with delay bounding techniques, iterative optimization, and further sequential performance tuning, which are
 numerically shown to provide nearly optimal performance. The contributions of this work are summarized below:
\begin{list}{$\bullet$}{\setlength{\leftmargin}{1em}\setlength{\labelsep}{.5em}
\setlength{\itemindent}{-.5em}}
\item
\emph{Conventional MCC with a non-computing AP}: We first consider
the conventional MCC with a non-computing AP, and formulate the
problem to jointly optimize the offloading decision and the
communication resource allocation of all tasks, to minimize a
weighted sum of the costs of energy, computation, and delay for all
users. The resulting mixed integer programming problem can be
reformulated as a non-convex quadratically constrained quadratic
program (QCQP) \cite{boyd2004}, which is NP-hard in general. To
solve this challenging problem, we first present a performance
bounding framework that utilizes both the upper and lower bounds of
the multi-task total communication and computation delay for each
user. We then propose an efficient Multi-User Multi-Task Offloading
(MUMTO) algorithm based on separable semidefinite relaxation (SDR)
\cite{luo2010}, with recovery of the binary offloading decision and
subsequent optimal allocation of the communication resource.

\item
\emph{MCC with a CAP}: We next consider the presence of a CAP in the
MCC, aiming to jointly optimize the task offloading decisions and
the allocation of computation and communication resources of all
tasks. However, the availability of CAP computation further
complicates mobile task offloading decisions, adding an extra
dimension of variability at the CAP. To solve this challenging
problem, we further propose an efficient three-step algorithm named
MUMTO with CAP (MUMTO-C), which first utilizes a generalized version
of the MUMTO SDR with an added CAP, and then performs additional
alternating optimization and sequential tuning. We show that it
always computes a locally optimal solution, which contains the
binary offloading decision and subsequent optimal allocation of the
computation and communication resources.

\item
\emph{Lower bounds and performance}: For both two scenarios
considered above, we obtain lower bounds for the minimum cost  as
the benchmark for performance evaluation. Simulation results show
that MUMTO and MUMTO-C both
 give nearly optimal performance under various parameter
settings. Furthermore, for the case with a CAP, we conduct simulation
experiments on alternative combinations of the three components of
the MUMTO-C algorithm, clarifying their roles and contributions to
the overall system performance. Finally, we compare the performance
of MUMTO-C against that of purely local processing, purely cloud
processing, and hybrid local-cloud processing without the CAP, which
demonstrates the effectiveness of the proposed algorithm in joint
management of the computation and communication resources in the
three-tier computing system of local devices, CAP, and remote cloud
server.
\end{list}

\emph{Organization}: The rest of this paper is organized as follows. In Section II, we discuss the related work. In Section III, we
describe the system model. In Section IV, we provide details of the
problem formulation, the proposed algorithm, and the lower bound of minimum system cost
for the conventional MCC without a CAP. In Section V, the impact on the presence of the CAP is
further studied.
We present numerical
results in Section VI and conclude in Section VII.

\emph{Notations}: Trace and transpose of matrix $\mathbf{A}$ are denoted by  $\mathrm{Tr}(\mathbf{A})$ and $\mathbf{A}^T$, respectively.  A positive semi-definite matrix   $\mathbf{A}$ is denoted as
$\mathbf{A} \succcurlyeq 0$.  Notation $\mathrm{diag}(\mathbf{a})$ denotes a
diagonal matrix with diagonal elements being elements of vector
$\mathbf{a}$, and   $\mathbf{A}(i,j)$ denotes the
$(i,j)\mathrm{th}$ entry of matrix $\mathbf{A}$.

\section{Related Work}
\subsection{Two-Tier Offloading System}
Many existing studies focus on two-tier cloud
networks with mobile users and another tier of external processors.

For a single user offloading its entire application, the tradeoff
between energy saving and computing performance was studied in
\cite{kumar2010,wen2012,munoz2015}. Different from the above
whole-application offloading, the authors of \cite{cuervo2010,
chun2011,kosta2012,zhang2013infocom,mahmoodi2016,kao2015,wu2016}
considered partitioning an application into multiple tasks.
Specifically, the authors of \cite{cuervo2010,chun2011,kosta2012} focus on the
implementation of offloading mechanisms from the mobile device to
the cloud, while the discussion on optimizing the offloading
decisions was limited. In \cite{zhang2013infocom}, a heuristic
offloading policy was proposed for a mobile user with sequential
tasks.
 In \cite{mahmoodi2016,kao2015,wu2016}, the problem of cloud offloading for a mobile user
with dependent tasks was studied.
All of the above studies focus on the single-user case. 

The case of task offloading by multiple mobile users has been
considered in
\cite{kaewpuang2013,ren2013,chen2014,meskar2017,chen2015efficient,sardellitti2015,lyu2017},
but in all of these works, each user only has a single task to
process.
Without considering resource allocation, the authors of
\cite{kaewpuang2013,chen2014,meskar2017,chen2015efficient} proposed
different approaches to obtain the offloading decisions for each
user.
In \cite{ren2013,sardellitti2015}, when all tasks are always
offloaded, the authors optimized the allocation of computation and
communication resources. In contrast to the above studies, instead
of optimizing either the offloading decision only or the resource
allocation only, in this work we study the joint optimization as
they are inter-dependent. The authors of \cite{lyu2017} considered
the joint allocation of offloading decision and resource allocation
with a sequential optimization heuristic. The method can only be
applied to the case where each user has a single task. In contrast,
in this work, the system design is much more challenging since we
consider the general multi-user multi-task scenario.

\subsection{Three-Tier Offloading System }
Besides the two-tier cloud networks above, the
three-tier network model, consisting of mobile users, a local
computing node (e.g., cloudlet or CAP), and a remote cloud server,
has been studied in
\cite{Rahimi2012,Rahimi2013,Song2014,cardellini2016,chen2015spawc,chen2016icassp,chen2018tmc}.
Compared with two-tier systems, the three-tier system adds extra
flexibility for task offloading.
In \cite{Rahimi2012,Rahimi2013,Song2014,cardellini2016}, the authors
 only focused on optimizing the offloading decisions without
considering the allocation of computation and communication
resources. However, since both computation and communication
resources are limited and shared among all users, without
efficiently allocating those limited resources to different users,
the full benefit of task offloading cannot be realized. The joint
optimization of the offloading decision and the allocation of
computation and communication resources for a general three-tier
multi-user multi-task offloading system has not been investigated
before, and it is much more complicated to solve.

We have previously studied the scheduling of computation and
communication resources in a CAP for \textit{a single mobile user}
\cite{chen2015spawc} and multiple mobile users each with \textit{a
single task} only \cite{chen2016icassp,chen2018tmc}, showing
substantial system performance improvement under such simplified
system models. In this work, we focus on the joint optimization
problem for a general \textit{multi-user multi-task} scenario.


\section{System Model}
\begin{figure}[t]
\centering
\includegraphics [scale =0.3]{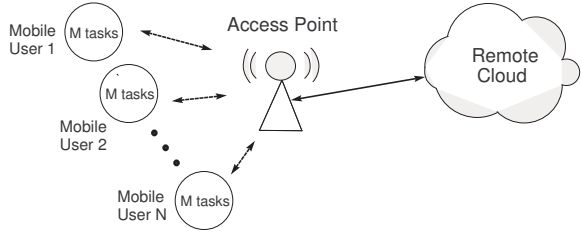}
\caption{Multi-user multi-task offloading system model. The AP may serve its conventional networking
function and forward tasks to the remote cloud server, or be a CAP
with built-in computation capability to directly process some
received tasks by itself. }
\label{fig:system_model}
\end{figure}

\subsection{Mobile Cloud Offloading with Multiple Users and Tasks}
 Consider a general cloud access network consisting
of one cloud server, one AP, and $N$ mobile users, each having $M$
independent tasks, as shown in
Fig.~\ref{fig:system_model}.\footnote{We assume the same number of tasks $M$ for all
users only for the notation simplicity. Our proposed solutions can be easily
extended to the general scenario where each mobile user has a different number
of tasks $M_i$.}
Examples of the AP may be a cellular base station or a WiFi access
point. The AP may serve its conventional networking function and
forward received tasks to the remote cloud server, or directly
process some of the tasks by itself when it has built-in computation
capability. In the latter case, we name it CAP. In this work, we
first study a conventional mobile cloud offloading scenario without
the CAP, aiming to obtain optimal offloading decisions for all
mobile users' tasks as well as resource allocation. Then, we will
further study a more general scenario with the presence of the CAP,
showing substantial system performance improvement. Notice that
we do not consider any specific queueing model for each user's tasks.
We will show latter in Sections \ref{sec.MUMTO} and
\ref{sec.MUMTO-C} that our proposed solutions are generally applicable to any queueing discipline.

\emph{Remark}: Our system model is a common one considered in many
previous studies
\cite{zhang2013infocom,kao2015,wu2016,kaewpuang2013,Song2014,sardellitti2015,chen2014,meskar2017,chen2015efficient,lyu2017},
where all $M$ tasks of each user are assumed to be available at the
starting time. For a dynamic system where the tasks arrive at
different times, we may apply our model and the proposed solution in
a quasi-static manner, where the system processes the tasks in
batches as they are collected \cite{Shmoys1995}. Also, note that for
the mobile cloud system considered, we  focus   on the bandwidth
sharing of wireless communication links among users, while assuming
that the statistics of each wireless link remain unchanged  during
the  processing of the users' tasks. This reflects a relatively
static or low-mobility scenario. The mobility issue and its effect
on the  offloading performance  is not considered this work and will
be left for future work.

The main symbols used in the system model are summarized in
Table~\ref{table_notation}.

\begin{table}[t]
\caption{List of Main Symbols} 
\centering
\small
\begin{tabular}{r| l}
\hline\hline 
\textbf{Symbol}& \textbf{Description} \\
\hline 
$E^l_{ij}$ & local processing energy of user $i$'s task $j$\\
$E^t_{ij}$,\ $E^r_{ij}$ & uplink transmitting energy and downlink\\
         & receiving energy of user $i$'s task $j$ between the\\
         & mobile user and the AP\\
$T^l_{ij}$, $T^a_{ij}$, $T^c_{ij}$ & local processing time, CAP processing time, and \\
         &cloud processing time of user $i$'s task $j$\\
$T^t_{ij}$,\ $T^r_{ij}$ & uplink transmission time and downlink \\
         & transmission time of user $i$'s task $j$ between the\\
         & mobile user and the AP\\
$T^{ac}_{ij}$ & transmission time of user $i$'s task $j$ between the\\
          & AP and the cloud\\
$C_{\textrm{UL}}$,\ $C_{\textrm{DL}}$ & uplink bandwidth and downlink bandwidth for \\
          & transmission between mobile users and the AP\\
$C_{\textrm{Total}}$ & total transmission bandwidth between mobile \\
          & users and the AP\\
$c^u_{i}$,\ $c^d_{i}$ & uplink bandwidth and downlink bandwidth \\
          & assigned to user $i$\\
$\eta_i^u$,\ $\eta_i^d$ & spectral efficiency of uplink and downlink \\
          &transmission between user $i$ and the AP\\
$r^{ac}$ & transmission rate between the AP and the cloud \\
$f^{a}_i$ & CAP processing rate assigned to user $i$'s tasks\\
$f_A$ &  total CAP processing rate\\
$f^{c}$ & cloud processing rate for each user\\
$C^a_{ij}$ & CAP usage cost of user $i$'s task $j$\\
$C^c_{ij}$ & cloud usage cost of user $i$'s task $j$\\
 \hline\hline
\end{tabular}
\label{table_notation}
\end{table}

\subsection{Cost of Local Processing}
We denote  by $D_{{\textrm{in}}}(ij)$,
$D_{{\textrm{out}}}(ij)$, and $Y(ij)$ the input data size, output
data size, and processing cycles\footnote{The processing cycles of
user $i$'s task $j$ depends on the input data size and the
application type.} of user $i$'s task $j$,
respectively.\footnote{These quantities may be obtained by applying a program profiler  \cite{cuervo2010, chun2011,kosta2012}, as similarly used in
\cite{wen2012,munoz2015,sardellitti2015,chen2014,meskar2017,chen2015efficient,lyu2017}.}
For task $j$ being locally processed by user $i$, the
corresponding energy consumed for processing is denoted by
$E^l_{ij}$ and the processing time is denoted by $T^l_{ij}$.

\subsection{Cost of Remote Cloud Processing}
When user $i$'s task $j$ is offloaded to the AP, we
denote by $E^t_{ij}$ and $E^r_{ij}$, respectively, the energy
consumed for wireless transmission and reception by the user. For
the wireless connections between mobile users and the AP, we denote
the uplink and downlink transmission times by
$T^t_{ij}=D_{{\textrm{in}}}(ij)/(\eta^u_{i}c^u_{i})$ and
$T^r_{ij}=D_{{\textrm{out}}}(ij)/(\eta^d_{i}c^d_{i})$, respectively,
where $c^u_{i}$ and $c^d_{i}$ are uplink and downlink bandwidth
allocated to user $i$, and $\eta_i^u$ and $\eta_i^d$ are the
spectral efficiency of uplink and downlink transmission between user
$i$ and the AP, respectively\footnote{The spectral efficiency can be
approximated by $\mathrm{log}(1+\mathrm{SNR})$ where $\mathrm{SNR}$
is the link quality between user $i$ and the AP.}. We have the
following constraints on $c^u_{i}$ and $c^d_{i}$ as they are limited
by the uplink bandwidth $C_{\textrm{UL}}$ and downlink bandwidth
$C_{\textrm{DL}}$
\begin{align}\label{a}
\sum_{i=1}^Nc^u_{i}\leq C_{\mathrm{UL}},
\end{align}
and
\begin{align}\label{b}
\sum_{i=1}^Nc^d_{i}\leq C_{\mathrm{DL}}.
\end{align}
We may consider also a total bandwidth constraint
\begin{align}\label{cap_total}
\sum_{i=1}^N(c^u_{i}+c^d_{i})\leq C_{\mathrm{Total}}.
\end{align}

Since the AP has to further offload the task to the cloud, there is the
additional transmission time between the AP and the cloud denoted
by
$T^{ac}_{ij}=(D_{{\textrm{in}}}(ij)+D_{{\textrm{out}}}(ij))/r^{ac}$,
and the cloud processing time denoted by $T^{c}_{ij}=Y(ij)/f^{c}$,
where
 $r^{ac}$ is the transmission
rate between the AP and the cloud and $f^c$ is the cloud processing
rate \textit{for each user}. The rate $r^{ac}$ is assumed to be a
pre-determined value regardless of the number of users, since the
AP-cloud link is likely to be a high-capacity wired connection in
comparison with the limited wireless links between the mobile users
and the AP, so that there is no need to consider bandwidth sharing
among the users. Similarly, $f^c$ is also assumed to be a
pre-determined value because of the high computational capacity and
dedicated service of the remote cloud server. Thus, $T^{ac}_{ij}$
and $T^{c}_{ij}$ only depend on user $i$'s task $j$ itself.

Finally,
the cloud usage cost of processing user $i$'s task $j$ at the cloud
is denoted by $C^c_{ij}$. The usage cost may depend on the data size
and processing cycles of a task and the hardware and energy cost to
maintain the cloud server, but such detail is outside the scope of this work.
Here we simply assume that $C^c_{ij}$ is given for all $i$ and $j$.

\subsection{Cost of CAP Processing}
When we consider the presence of a CAP, some of the offloaded tasks can be directly
processed by the CAP. If user $i$'s task $j$ is processed by the
CAP (i.e., instead of being further forwarded to the remote cloud),
besides the communication energy (i.e., $E^t_{ij}$ and $E^r_{ij}$)
and delay (i.e., $T^t_{ij}$ and $T^r_{ij}$) mentioned above, we
denote the CAP processing time by $T^a_{ij}=Y(ij)/f^a_i$, where
$f^a_i$ is the assigned processing rate, which is limited by the
total processing rate $f_A$ at the CAP
\begin{align}\label{problem_CAP.a}
\sum_{i=1}^Nf^a_{i}\leq f_{A}.
\end{align}
Similarly, denote the CAP usage cost of processing user $i$'s task
$j$ at the CAP by $C^a_{ij}$. In the following, we first study the
conventional MCC without considering the CAP. The impact on the
presence of a CAP will be further studied in Section \ref{sec_CAP}.

\section{Multi-user Multi-task Offloading without CAP}
In this section, we study the conventional mobile cloud network
where the AP always forwards the received tasks to the remote cloud
server. In this case, we have a two-tier offloading system, and we
focus on jointly optimizing the offloading decision and the
communication resource allocation of all tasks, to minimize a
weighted sum of the costs of energy, computation, and the delay for
all users. 
\subsection{Offloading Decision}
Since there is no CAP, each mobile user can either process its tasks locally or offload
some of them to the cloud for processing through the AP. Let
$x_{ij}$ denote the offloading decision for task $j$ of  user $i$,
given by
\begin{equation*}
x_{ij} =
\begin{cases}
0, & \text{process task $j$ of user $i$ locally;}\\
1, & \text{offload task $j$ of user $i$  to the cloud}.
\end{cases}
\end{equation*}

\subsection{Problem Formulation}\label{sec.mcc_prob}

We aim at reducing mobile users' energy consumption and maintain the
service quality of processing their tasks, measured by the delays
incurred due to transmission and/or processing. For this goal, we
define the total system cost as the weighted sum of total energy
consumption, the costs to offload and process all tasks, and the
corresponding transmission and processing delays for all users. Our
objective is to minimize the total system cost by jointly optimizing
the task offloading decisions $x_{ij}$ and the communication
bandwidth resource allocation $\rbf_i=[c^u_{i},\ c^d_{i}]^T$.
This optimization problem is formulated as follows:
\begin{align}
\min\limits_{\{x_{ij}\!\},\{\rbf_i\!\}}
&\quad \!\!\!\!\!\sum_{i=1}^N\!\bigg[\!\sum_{j=1}^M(E^l_{ij}(1-x_{ij})\!+E^C_{ij}x_{ij})\! +\rho_i\max\{T^L_{i}, T^C_{i}\}\!\bigg]\label{prob_noCAP}\allowdisplaybreaks\\
\text{s.t.}
&\quad \!\!\!\eqref{a},\eqref{b},\eqref{cap_total},\allowdisplaybreaks\nonumber\\
&\quad \!\!\!r^u_{i}, r^d_{i},\geq 0,\forall i,\label{c}\allowdisplaybreaks\\
&\quad \!\!\!x_{ij}\in \{0,1\},\forall i, j,\label{d}
\end{align}
where $E^C_{ij}\triangleq(E^t_{ij}+E^r_{ij}+\beta C^c_{ij})$ is the
weighted transmission energy and processing cost of offloading and
processing task $j$ of user $i$ to the cloud, with $\beta$ being the
relative weight; in addition, $T^L_{i}$ is the processing delay of
tasks processed by the mobile user $i$ itself,
$T^C_{i}$ is the overall transmission and remote-processing delay
for tasks of mobile user $i$ processed at the cloud, and $\rho_i$ is
the weight on the task processing delay relative to energy
consumption in the total system cost.

Depending on the performance
requirement, the value of $\rho_i$ can be adjusted to impose
different emphasis on delay and energy consumption.\footnote{ To avoid
mathematical redundancy, we only put the weight in front of the
delay and normalize the weighted sum cost to have the unit of
energy. However, it can be easily extended to an objective with some
arbitrary unit (e.g., dollars).} The proposed optimization problem
\eqref{prob_noCAP} can be solved by any controller in this network
after collecting all required information. In practice, the
controller could be the AP. That is, each user provides its
information to the AP, and the AP broadcasts the obtained
offloading decisions (and the corresponding resource allocations) to
all users by solving problem \eqref{prob_noCAP}.

The above mixed-integer programming problem is difficult to solve in
general. Based on the offloading decision $x_{ij}$ for each task, we
have the total local processing delay for each user $T^L_{i}= \sum_{j=1}^MT^l_{ij}(1-x_{ij})$, for all $i$.
However, we note that the overall delay for remote processing,
$T^C_i$, is challenging to calculate exactly. This is because, when
there are multiple tasks offloaded by a users, the transmission
times and processing times may overlap in an unpredictable manner,
which depends on the offloading decision, communication resource
allocation, and task scheduling order. In fact, since $T^C_i$
consists of the uplink transmission times, remote-processing time,
and downlink transmissions times of all tasks, it may be viewed as
the output of a multi-machine flowshop schedule, which remains an
open research problem \cite{Garey1976}. Since $T^C_i$ is not
precisely tractable, we will use both upper and lower bounds of $T^C_i$ in our proposed solution and performance benchmarking.
Under the MUMTO algorithm, they are shown to give total system costs that are close to each
other.

\subsection{Multi-user Multi-task
Offloading (MUMTO) Algorithm}\label{sec.MUMTO}
The joint optimization problem \eqref{prob_noCAP} is a mixed-integer
non-convex programming problem. To find an efficient solution to the
original problem \eqref{prob_noCAP}, in the following, we first
propose both upper bound and lower bound formulations of
$T^{C}_{i}$, then transform the optimization problem
\eqref{prob_noCAP} into a separable QCQP, and finally propose a
separable SDR approach to obtain the binary offloading decisions
$\{x_{ij}\}$ and the communication resource allocation $\{\rbf_i\}$.
\subsubsection{\textbf{Bounds of Remote-Processing Delay}}\label{bound_of_Tc_noCAP}
When a mobile user offloads more than one task to the cloud, there
will be overlaps in the communication and processing times as
mentioned above, making it difficult to exactly characterize the
overall delay $T^{C}_{i}$. However, we have the following upper bound of
$T^C_{i}$ as the \textit{worst-case delay} formulation:
\begin{align} \label{worst_Tc}\hspace*{-0.1cm}
T^{C_{(U)}}_{i}\hspace*{-0.05cm}=\hspace*{-0.05cm}\sum_{j=1}^M\hspace*{-0.05cm}\left(\hspace*{-0.05cm}\frac{D_{{\textrm{in}}}(ij)}{\eta^u_ic^u_i}+\frac{D_{{\textrm{out}}}(ij)}{\eta^d_ic^d_i}+T^{ac}_{ij}+T^{c}_{ij}\right)x_{ij},\
\forall i.
\end{align}
Since the worst-case delay sums the transmission delays and
processing delays together without any overlap, it will always be
greater than the real delay given the same offloading decision and
resource allocation. On the other hand, we separate the offloading
delays of all mobile users into several components and only consider
the largest one as the lower bound of $T^C_{i}$:
\begin{align}\label{best_Tc}
T^{C_{(L)}}_{i}=\max\{T^u_{i}, T^d_{i}, T^{uac}_{i}, T^{dac}_{i},
T^{c'}_{i}\},\ \forall i,
\end{align}
where
$T^u_{i}=\sum_{j=1}^MD_{{\textrm{in}}}(ij)x_{ij}/(\eta^u_ic^u_i)$
and
 $T^d_{i}= \sum_{j=1}^MD_{{\textrm{out}}}(ij)x_{ij}/(\eta^d_ic^d_i)$ are total
 uplink and downlink transmission times between the user and the AP
 for user $i$, respectively, $T^{uac}_{i}=\sum_{j=1}^MD_{{\textrm{in}}}(ij)x_{ij}/r^{ac}$
 and $T^{dac}_{i}= \sum_{j=1}^MD_{{\textrm{out}}}(ij)x_{ij}/r^{ac}$
 are total uplink and downlink transmission times between the AP and
 the cloud for user $i$, respectively, and $T^{c'}_{i}=\sum_{j=1}^MY(ij)x_{ij}/f^{c}$
 is the total cloud processing time for user $i$.

In the following, we will use the worst-case delay $T^{C_{(U)}}_{i}$
in optimization problem \eqref{prob_noCAP} to obtain an approximate
solution, which can provide an upper bound to the total system cost.
We then use $T^{C_{(L)}}_{i}$ similarly, to obtain a lower bound of
the total system cost, for performance benchmarking. In Section
\ref{sec_simulation}, by comparing both cases, we show that the
MUMTO algorithm based on the worst case formulation gives nearly
optimal performance.

\subsubsection{\textbf{QCQP Transformation and Semidefinite Relaxation}}\label{sec_QCQP_noCAP}
We first replace $T^{C}_{i}$ with $T^{C_{(U)}}_{i}$ in problem
\eqref{prob_noCAP}, and rewrite the integer constraint \eqref{d} as
\begin{align} \label{xy_new}
x_{ij}(x_{ij}-1)=0,\ \forall i, j.
\end{align}
We also introduce a additional auxiliary variable $t_i$ for
$\max\{T^L_{i}, T^{C_{(U)}}_{i}\}$, the problem \eqref{prob_noCAP}
is now transformed into the following equivalent problem:
\begin{align}
\hspace*{-.15cm}
\min\limits_{\{x_{ij}\},\{\rbf_i,t_i\}}\hspace*{-.1cm}
&\ \sum_{i=1}^N\bigg[\sum_{j=1}^M(E^l_{ij}(1-x_{ij})+E^C_{ij}x_{ij})+\rho_it_i\bigg]\label{prob_noCAP_new}\allowdisplaybreaks\\
\text{s.t.} &\  \sum_{j=1}^MT^l_{ij}(1-x_{ij})\leq t_i,\ \forall i,\label{prob_noCAP_new.a}\\
&\
\sum_{j=1}^M\hspace*{-.05cm}\left(\hspace*{-.05cm}\frac{D_{{\textrm{in}}}(ij)}{\eta^u_ic^u_i}\hspace*{-.05cm}+\hspace*{-.05cm}
\frac{D_{{\textrm{out}}}(ij)}{\eta^u_ic^d_i}\hspace*{-.05cm}+\hspace*{-.05cm}T^{ac}_{ij}\hspace*{-.05cm}+\hspace*{-.05cm}T^{c}_{ij}\hspace*{-.05cm}\right)\hspace*{-.05cm}x_{ij}\hspace*{-.05cm}\leq\hspace*{-.05cm} t_i,\forall i,\label{prob_noCAP_new.b}\\
&\quad  \eqref{a},\eqref{b},\eqref{cap_total},\eqref{c}, \text{~and} \ \eqref{xy_new}. \nonumber
\end{align}
In order to obtain the eventual SDR formulation, we first transform
the optimization problem \eqref{prob_noCAP_new} into a separable
QCQP problem by the following steps.

First, we introduce two auxiliary variables
$D^u_{i}$ and $D^d_{i}$, and replace  constraint \eqref{prob_noCAP_new.b}
 with the following equivalent constraints:
\begin{align}\label{constraint_delay}
&D^u_{i}+D^d_{i}+\sum_{j=1}^M(T^{ac}_{ij}+T^{c}_{ij})x_{ij}\leq t_i,\ \forall i,
\end{align}
\begin{align}\label{constraint_D^u}
\sum_{j=1}^M\frac{D_{{\textrm{in}}}(ij)x_{ij}}{\eta^u_ic^u_i}\leq
D^u_{i},\ \forall i,
\end{align}
and
\begin{align}\label{constraint_D^d}
\sum_{j=1}^M\frac{D_{{\textrm{out}}}(ij)x_{ij}}{\eta^d_ic^d_i}\leq
D^d_{i},\ \forall i,
\end{align}
where \eqref{constraint_delay} is the overall offloading delay
constraint, and \eqref{constraint_D^u} and \eqref{constraint_D^d}
correspond to the uplink transmission time and the downlink
transmission time, respectively.

Next, we vectorize the
variables and parameters in problem \eqref{prob_noCAP_new}. Define
\begin{align}\label{def_w} \wbf_i \triangleq[x_{i1},\ldots,
x_{iM},c^u_{i},D^u_{i},c^d_{i},D^d_{i},t_i]^T,\ \forall i,
\end{align}
which is the decision vector for user $i$ with all decision
variables. Then, the objective in problem \eqref{prob_noCAP_new} can
be rewritten as
\begin{align}
\sum_{i=1}^N\mathbf{b}_i^T\wbf_i+\sum_{i=1}^N\sum_{j=1}^ME^l_{ij},\label{problem_QCQP_noCAP}
\end{align}
where $\mathbf{b}_{i}\triangleq[(E^C_{i1}-E^l_{i1}),\ldots, \ (E^C_{iM}-E^l_{iM}),\ \mathbf{0}_{1\times 4},\ \rho_i]^T$.
We rewrite the local processing delay constraint
\eqref{prob_noCAP_new.a} as
\begin{align}
({\mathbf{b}^l_i})^T\wbf_i\leq -\sum_{j=1}^MT^l_{ij},\ \forall i,\label{problem_QCQP_noCAP.a}
\end{align}
where $\mathbf{b}^l_i\triangleq-[T^l_{i1},\ldots, \ T^l_{iM},\ \mathbf{0}_{1\times 4},\ 1]^T$.
For the cloud processing delay constraint \eqref{constraint_delay}, it can be rewritten as
\begin{align}
({\mathbf{b}^c_i})^T\wbf_i\leq 0,\ \forall i, \label{problem_QCQP_noCAP.b}
\end{align}
where $\mathbf{b}^c_i\triangleq[(T^{ac}_{i1}+T^c_{i1}),\ldots, \ (T^{ac}_{iM}+T^c_{iM}),0,1,0,1,-1]^T$.
The matrix forms of constraints \eqref{constraint_D^u} and \eqref{constraint_D^d} are
\begin{align}
\mathbf{w}_i^T\mathbf{A}^\mu_{i}\mathbf{w}_i+({\mathbf{b}^\mu_{i}})^T\mathbf{w}_i\leq 0,\ \  \mu\in\{u,d\},\ \forall i,\label{problem_QCQP_noCAP.c}\allowdisplaybreaks
\end{align}
where
\begin{align}
&\mathbf{A}^{\mu'}_i\triangleq -\frac{1}{2}
\begin{bmatrix}
0&\eta^{\mu}_i\\
\eta^{\mu}_i&0\\
\end{bmatrix},\ \ \mu\in\{u,d\},\nonumber\allowdisplaybreaks\\
&\mathbf{A}^u_i\triangleq
\begin{bmatrix}
\mathbf{0}_{M\times M}&\mathbf{0}_{M\times 2}&\mathbf{0}_{M\times 3}\\
\mathbf{0}_{2\times M}&\mathbf{A}^{u'}_i&\mathbf{0}_{2\times 3}\\
\mathbf{0}_{3\times M}&\mathbf{0}_{3\times 2}&\mathbf{0}_{3\times 3}\\
\end{bmatrix},\nonumber\allowdisplaybreaks\\
&\mathbf{A}^d_i\triangleq
\begin{bmatrix}
\mathbf{0}_{(M+2)\times (M+2)}&\mathbf{0}_{(M+2)\times 2}&\mathbf{0}_{(M+2)\times 1}\\
\mathbf{0}_{2\times (M+2)}&\mathbf{A}^{d'}_i&\mathbf{0}_{2\times 1}\\
\mathbf{0}_{1\times (M+2)}&\mathbf{0}_{1\times 2}&0\\
\end{bmatrix},\nonumber\allowdisplaybreaks\\
&\mathbf{b}^u_i\triangleq[\mathbf{D}_{\textrm{in}}(i1),\ldots, \
\mathbf{D}_{\textrm{in}}(iM),\ \mathbf{0}_{1\times 5}]^T, \nonumber\allowdisplaybreaks\\
&\mathbf{b}^d_i\triangleq[\mathbf{D}_{\textrm{out}}(i1),\ldots, \
\mathbf{D}_{\textrm{out}}(iM),\ \mathbf{0}_{1\times
5}]^T.\nonumber\allowdisplaybreaks
\end{align}

The uplink and downlink bandwidth resource constraints \eqref{a} and
\eqref{b} correspond to
\begin{align}
\sum_{i=1}^N({\mathbf{b}^{U}_i})^T\wbf_i= C_{\mathrm{UL}},\label{problem_QCQP_noCAP.d}\allowdisplaybreaks
\end{align}
and
\begin{align}
\sum_{i=1}^N({\mathbf{b}^{D}_i})^T\wbf_i= C_{\mathrm{DL}},\label{problem_QCQP_noCAP.e}\allowdisplaybreaks
\end{align}
respectively, where $\mathbf{b}^U_i\triangleq[\mathbf{0}_{1\times
M},\ 1,\ \mathbf{0}_{1\times 4}]^T$ and
$\mathbf{b}^D_i\triangleq[\mathbf{0}_{1\times M+2},\ 1,\
\mathbf{0}_{1\times 2}]^T.$
Similarly, the total bandwidth constraint \eqref{cap_total} is
rewritten as
\begin{align}
\sum_{i=1}^N({\mathbf{b}^{S}_{i}})^T\mathbf{w}_i\leq
C_{\mathrm{Total}},\label{problem_QCQP.total}\allowdisplaybreaks
\end{align}
where $\mathbf{b}^{S}_i\triangleq[\mathbf{0}_{1\times
M},1,0,1,0,0]^T.$
The constraint \eqref{c} used to ensure all variables great than or equal to $0$ is replaced by
\begin{align}
\wbf_i \succcurlyeq 0,\ \forall i.\label{problem_QCQP_noCAP.f}\allowdisplaybreaks
\end{align}
Finally, we rewrite the integer constraint \eqref{xy_new} as
\begin{align}
\wbf_i^T\textrm{diag}(\ebf_j)\wbf_i-\ebf_j^T\wbf_i=0,\ \forall i,j,\label{problem_QCQP_noCAP.g}\allowdisplaybreaks
\end{align}
where $\ebf_{j}$ as the $(M+5)\times 1$ standard unit vector with
the $j$th entry being $1$.

By further defining $\zbf_i \triangleq[\wbf_i^T, 1]^T$, together
with the above matrix form presentations, and dropping the constant term
$\sum_{i=1}^N\sum_{j=1}^ME^l_{ij}$ from the objective function in
\eqref{problem_QCQP_noCAP}, problem
\eqref{prob_noCAP_new} can now be further transformed into the
following homogeneous separable QCQP formulation:
\begin{align}
\min\limits_{\{\zbf_i\}}
&\quad \sum_{i=1}^N\zbf_i^T\Gbf_i\zbf_i\label{Eq_homo_QCQP}\allowdisplaybreaks\\
\text{s.t.}
&\quad \zbf_i^T\Gbf^l_{i}\zbf_i\leq -\sum_{j=1}^MT^l_{ij},\ \forall i,\allowdisplaybreaks\label{Eq_homo_QCQP.a}\\
&\quad \zbf_i^T\Gbf^c_{i}\zbf_i\leq 0,\ \forall i,\allowdisplaybreaks\label{Eq_homo_QCQP.b}\\
&\quad \zbf_i^T\Gbf^\mu_{i}\zbf_i\leq 0,\ \mu\in\{u,d\},\ \forall i,\allowdisplaybreaks\label{Eq_homo_QCQP.c}\\
&\quad \sum_{i=1}^N\zbf_i^T\Gbf^U_i\zbf_i\leq C_{\mathrm{UL}},\ \sum_{i=1}^N\zbf_i^T\Gbf^D_i\zbf_i\leq C_{\mathrm{DL}},\label{Eq_homo_QCQP.d}\allowdisplaybreaks\\
&\quad \sum_{i=1}^N\zbf_i^T\Gbf^S_i\zbf_i\leq C_{\mathrm{Total}},\allowdisplaybreaks\label{Eq_homo_QCQP.cap_total}\\
&\quad \zbf_i^T\Gbf^I_j\zbf_i=0,\ \forall i,j,\label{Eq_homo_QCQP.e}\\
&\quad \zbf_i \succcurlyeq 0,\ \forall
i,\allowdisplaybreaks\label{Eq_homo_QCQP.f}
\end{align}
where
\begin{align}
&\Gbf_i \triangleq
\begin{bmatrix}
\mathbf{0}& \frac{1}{2}\mathbf{b}_i\\
\frac{1}{2}\mathbf{b}_i^T&0\\
\end{bmatrix},\ \
\Gbf^\mu_i\triangleq
\begin{bmatrix}
\mathbf{A}^\mu_i&\frac{1}{2}\mathbf{b}^\mu_i\\
\frac{1}{2}({\mathbf{b}^\mu_i})^T&0\\
\end{bmatrix},\ \mu\in\{u,d\},\allowdisplaybreaks\nonumber\\
&\Gbf^\pi_i\triangleq
\begin{bmatrix}
\mathbf{0}&\frac{1}{2}\mathbf{b}^\pi_i\\
\frac{1}{2}({\mathbf{b}^\pi_i})^T&0\\
\end{bmatrix},\ \pi\in\{l,c,U,D,S\},\nonumber\allowdisplaybreaks\\
&\Gbf^I_j\triangleq
\begin{bmatrix}
\textrm{diag}(\ebf_j)&-\frac{1}{2}\ebf_j\\
-\frac{1}{2}\ebf_j^T&0\\
\end{bmatrix}.\nonumber
\end{align}
As problems \eqref{prob_noCAP_new} and
\eqref{Eq_homo_QCQP} are equivalent, all constraints have one-to-one
correspondence. 

The optimization problem \eqref{Eq_homo_QCQP} is a non-convex
separable QCQP problem \cite{boyd2004}. This problem is NP-hard. To
show this, first, we note that problems \eqref{prob_noCAP_new} and
\eqref{Eq_homo_QCQP} are equivalent. For problem
\eqref{prob_noCAP_new}, when we only consider the offloading
decisions as variables (i.e., each user has already been assigned
some fixed communication and computation resources), the problem is
reduced to a linear integer programming problem. Then, if the $t_i$
values are further given, (e.g., $t_i = \sum_{j=1}^M T_{ij}$),
problem \eqref{prob_noCAP_new} is reduced to the 0-1 knapsack
problem, which is NP-hard.

To find an approximate solution,
we apply the separable SDR approach \cite{luo2010}, where we relax
the problem into a separable semidefinite programming (SDP) problem.
Specifically, define $\mathbf{Z}_i\triangleq\zbf_i\zbf_i^T$. The
following equality holds:
\begin{align}
\mathbf{z}_i^T\mathbf{G}\mathbf{z}_i = \textrm{Tr}(\mathbf{G}\mathbf{Z}_i),
\end{align}
with $\textrm{rank}(\mathbf{Z}_i)=1$. By dropping the rank
constraint $\textrm{rank}(\mathbf{Z}_i)=1$, we have the following
separable SDP problem: 
\allowdisplaybreaks
\begin{align}
\min\limits_{\{\mathbf{Z}_i\}}
&\quad \sum_{i=1}^N\textrm{Tr}(\Gbf_i\Zbf_i)\label{Eq_SDP_QCQP_noCAP}\allowdisplaybreaks\\
\text{s.t.}
&\quad \textrm{Tr}(\Gbf^l_{i}\Zbf_i)\leq -\sum_{j=1}^MT^l_{ij},\ \forall i,\allowdisplaybreaks\label{Eq_SDP_QCQP_noCAP.a}\\
&\quad \textrm{Tr}(\Gbf^r_{i}\Zbf_i)\leq 0,\  r\in\{c,u,d\},\ \forall i,\allowdisplaybreaks\label{Eq_SDP_QCQP_noCAP.b}\\
&\quad \sum_{i=1}^N\textrm{Tr}(\Gbf^U_i\Zbf_i)\leq C_{\mathrm{UL}},\ \sum_{i=1}^N\textrm{Tr}(\Gbf^D_i\Zbf_i)\leq C_{\mathrm{DL}},\label{Eq_SDP_QCQP_noCAP.c}\\
&\quad \sum_{i=1}^N\textrm{Tr}(\Gbf^S_i\Zbf_i)\leq C_{\mathrm{Total}},\label{Eq_SDP_QCQP_noCAP.cap_total}\\
&\quad \textrm{Tr}(\Gbf^I_j\Zbf_i)=0,\ \forall i,j,\allowdisplaybreaks\label{Eq_SDP_QCQP_noCAP.e}\\
&\quad \Zbf_i(M+6,M+6)=1,\ \forall i,\\
&\quad \Zbf_i\succcurlyeq 0,\ \forall i.\label{Eq_SDP_QCQP_noCAP.f}
\end{align}

%

The optimal solution $\{\Zbf_i^*\}$ to the above separable SDP
problem can be obtained efficiently in polynomial time using
standard SDP software, such as SeDuMi \cite{grant2009}. However,
since problem \eqref{Eq_SDP_QCQP_noCAP} is a relaxation of the problem
\eqref{Eq_homo_QCQP}, the optimal objective of the problem \eqref{Eq_SDP_QCQP_noCAP}
is only a lower bound of the optimal solution of the problem
\eqref{Eq_homo_QCQP} if $\{\Zbf_i^*\}$ does not have rank 1. Therefore, once
$\{\Zbf_i^*\}$  is obtained, we still need to recover a rank-1
solution from $\{\Zbf_i^*\}$ for the original problem \eqref{prob_noCAP}.
In the following, we propose an
algorithm to obtain the binary offloading decisions $\{x_{ij}\}$ and
the corresponding optimal communication resource allocation
$\{\rbf_i\}$ for problem \eqref{prob_noCAP}.

\subsubsection{\textbf{Binary Offloading Decisions and Resource
Allocation}}\label{sec.MDBSDR}

\begin{algorithm}[!t]
  \caption{MUMTO Algorithm}
  \begin{algorithmic}[1]
  \small
    \State Obtain optimal solution $\Zbf_i^*$'s of the separable SDP problem \eqref{Eq_SDP_QCQP_noCAP}.
    \State Extract $\Zbf^*_i(M+6,k)$, for $k=1,...,M$, from $\mathbf{Z}_i^*$.
 \State Record the values of $\Zbf^*_i(M+6,k)$, for $k=1,...,M$, as $\pbf_i=[p_{i1},\ldots,
 p_{iM}]^T$.
      \State Set $x^\mathrm{sdr}_{ij}=\mathrm{round}(p_{ij}),\forall i,j$.
      \State Set $\xbf^\mathrm{sdr}=[(\xbf^\mathrm{sdr}_1)^T,\ldots,
(\xbf^\mathrm{sdr}_N)^T]^T$, where
$\xbf^\mathrm{sdr}_i=[x^\mathrm{sdr}_{i1},\ldots,
x^\mathrm{sdr}_{iM}]^T$.
      \State Solve the resource allocation problem \eqref{prob_resource_noCAP} based on $\xbf^\mathrm{sdr}$;
      \State Compare the minimum cost of \eqref{prob_resource_noCAP} under $\xbf^{\mathrm{sdr}}$ with those under the local processing only and cloud processing only solutions. Select the one that yields the minimum system cost as  $\xbf^{\mathrm{sdr}^*}$.
    \State Output: the proposed offloading solution $\xbf^{\mathrm{sdr}^*}$  and the corresponding optimal resource
    allocation $\{\rbf_i^{\mathrm{sdr}^*}\}$.
  \end{algorithmic}
  \label{algorithm_noCAP}
\end{algorithm}

Define the offloading solution vector as $\xbf \triangleq
[\xbf_1^T,\ldots, \xbf_N^T]^T$, where $\xbf_i \triangleq
[x_{i1},\ldots, x_{iM}]^T$, for all $i$. Since the rank-1 constraint
has been removed from the relaxed problem \eqref{Eq_SDP_QCQP_noCAP},
the obtained solution $\Zbf_i^*$ for problem
\eqref{Eq_SDP_QCQP_noCAP} contains only real numbers. Our goal is to
obtain appropriate offloading decisions from $\Zbf_i^*$ by mapping
its elements to binary numbers.
 Note that only the first $M$ elements
in $\zbf_i$ correspond to the offloading decision variables for user
$i$ (see $\wbf_i$ in \eqref{def_w}). Also, we have
$\mathbf{Z}_i=\zbf_i\zbf_i^T$ and $\mathbf{z}_i(M+6)=1$, which means
the last row of $\mathbf{Z}_i$ satisfies $\Zbf_i(M+6,k)=\zbf_i(k)$,
for all $k$. Hence, we can use the values of $\Zbf^*_i(M+6,k)$ to
recover the binary offloading decision $\zbf_i(k)$, for $k=1,...,M$.
In addition, it can be shown that $\Zbf^*_i(M+6,k) \in [0,1]$, for
$k=1,...,M$. Define
$\pbf_i\triangleq[p_{i1},\ldots, p_{iM}]^T\triangleq [\Zbf^*_i(M+6,1),\cdots,\Zbf^*_i(M+6,M)]^T$. We have $p_{ij} \in [0,1]$,
 $\forall i,j$.  We recover the feasible decisions
$\xbf_i^{\mathrm{sdr}}$ using $\pbf_i$, where
$x_{ij}^{\mathrm{sdr}}=\mathrm{round}(p_{ij})$ is the rounding
result, and obtain the overall offloading decision as
$\xbf^\mathrm{sdr}=[(\xbf_1^\mathrm{sdr})^T,\ldots,
(\xbf_N^\mathrm{sdr})^T]^T$.

Once the offloading decision $\xbf^\mathrm{sdr}$ is obtained, the
optimization problem \eqref{prob_noCAP} reduces to the optimization
of communication resource allocation $\{\rbf_i\}$, which is given by
\begin{align}
\min\limits_{\{\rbf_i\}}
&\quad \bigg(E+\sum_{i=1}^N\rho_i\max\{T_{L_{i}}, T^{C_{(U)}}_{i}\}\bigg)\label{prob_resource_noCAP}\allowdisplaybreaks\\
\text{s.t.} &\quad \eqref{a},\eqref{b},\eqref{cap_total},
\text{~and} \ \eqref{c},\nonumber
\end{align}
where
$E\triangleq\sum_{i=1}^N\sum_{j=1}^M(E^l_{ij}(1-x_{ij})+E^C_{ij}x_{ij})$
is a constant value once  $\{x_{ij}\}$ are given. This resource
allocation problem \eqref{prob_resource_noCAP} is convex, which can
be solved optimally using standard convex optimization solvers. Note
that to obtain the best offloading decision, in practice, we should
compare $\xbf^\mathrm{sdr}$ with local processing only  and cloud
processing only decisions, and select the one resulting in the
minimum total system cost objective of \eqref{prob_resource_noCAP}
as the final offloading decision $\xbf^{\mathrm{sdr}^*}$.

We summarize MUMTO in Algorithm \ref{algorithm_noCAP}. Notice that
the SDP problem \eqref{Eq_SDP_QCQP_noCAP} can be solved within
precision $\epsilon$ by the interior point method in
$O(\sqrt{MN}\log(1/\epsilon))$ iterations, where the amount of work
per iteration is $O(M^{6}N^{4})$ \cite{nesterov1994}, while there
are $2^{MN}$ choices in exhaustive search to find the optimal
offloading decision. In addition, once the offloading decision is
made, we may schedule the multiple tasks to be offloaded in any
arbitrary order. The resultant $T^C_i$ will be less than
$T^{C_{(U)}}_{i}$.
 To measure
the effectiveness of this solution, in the following, we introduce a
lower bound of the optimal solution to the original problem
\eqref{prob_noCAP}.

\subsection{Lower Bound on the Optimal Solution}\label{sec_lowerbound_noCAP}
Previously, the cost function in our original optimization problem
\eqref{prob_noCAP} considers the worst-case
transmission-plus-processing delay \eqref{worst_Tc} for all users.
Once the offloading decision is made, we may schedule the multiple
tasks to be offloaded in any arbitrary order. The resultant
$T^{C}_i$ will be less than $T^{{C}_{(U)}}_i$. Therefore, the actual
cost based on MUMTO will be lower than the worst-case
cost.

However, we are still interested in the performance of
MUMTO compared with an optimal solution. Therefore, we
introduce a lower bound of the optimal solution to the original
problem \eqref{prob_noCAP}. We first introduce a new optimization
problem, where $T^{{C}_{(L)}}_i$ are used instead of  $T_i^{C}$ and
the objective function is replaced by its lower bound, as follows:
\begin{align}\label{Eq_new_objective_noCAP}
\hspace*{-1em}\min\limits_{\{x_{ij}\},\{\rbf_i\}}
&\quad \sum_{i=1}^N\bigg[\sum_{j=1}^M(E^l_{ij}(1-x_{ij})+E^C_{ij}x_{ij})\nonumber\\
&\quad \quad \ \ +\rho_i\max\{T^L_{i}, T^u_{i}, T^d_{i}, T^{uac}_{i}, T^{dac}_{i}, T^{c'}_{i}\}\bigg]\\
\text{s.t.} &\quad \eqref{a},\eqref{b},\eqref{cap_total},\eqref{c},
\text{~and} \ \eqref{d}.\nonumber
\end{align}
Notice that under the same offloading decisions and communication
resource allocation, this new objective function will always give us
a lower cost than the real cost.

Since the above optimization problem \eqref{Eq_new_objective_noCAP}
is still non-convex, we formulate a separable SDR problem similar to
\eqref{Eq_SDP_QCQP_noCAP}, whose details are omitted due to page
limitation. We note that the optimal objective of this SDR problem
is smaller than the optimal objective of
\eqref{Eq_new_objective_noCAP}. Hence, it can serve as a lower bound
of the minimum total system cost defined by the original
optimization problem \eqref{prob_noCAP}. In Section
\ref{sec_simulation_noCAP}, we show that MUMTO provides
nearly optimal performance under a wide range of parameter settings.
\section{Multi-user Multi-task Offloading with CAP}\label{sec_CAP}
When we consider the presence of a CAP, it may serve its conventional networking
function and forward the task to the remote cloud server, or
directly process the task by itself. Each task may be processed
locally at the mobile device, at the CAP, or at the remote cloud
server. An optimal offloading decision must take into consideration
the computation and communication energies, computation costs, and
communication and processing delays at all three locations.
In this
section, we further study the mobile cloud computing network with
the presence of the CAP, aiming to jointly optimize the task
offloading decisions and the communication and CAP processing
resource allocation.

\subsection{Offloading Decision}
Each mobile user can process its tasks locally or offload some of
them. With the presence of a CAP, those offloaded tasks may be
processed at the CAP or be further forwarded to the remote cloud.
Instead of only using $x_{ij}$, we denote the offloading decisions
for user $i$'s task $j$ by $x^l_{ij}, x^a_{ij}, x^c_{ij} \in
\{0,1\}$, indicating whether user $i$'s task $j$ is processed
locally, at the CAP, or at the cloud, respectively. The offloading
decisions are constrained by
\begin{equation}\label{eq_placement}
x^l_{ij} + x^a_{ij} + x^c_{ij} = 1.
\end{equation}
Notice that only one of $x^l_{ij}, x^a_{ij}$, and $x^c_{ij}$ for
user $i$'s task $j$ could be $1$.

\subsection{Problem Formulation}
The new total system cost is defined as the weighted sum of total
energy consumption, the costs to offload and process all tasks, and
the transmission and processing delays for all users. Define offloading decision vector $\xbf_{ij}\triangleq[x^l_{ij}, x^a_{ij}, x^c_{ij}]^T$. With a CAP,  both communication and CAP processing resources needs to be considered, defined by $\rbf_i \triangleq [c_i^u, c_i^d, f_i^a]^T$.
Similar to
Section \ref{sec.mcc_prob}, our objective is to minimize the total
system cost by jointly optimizing the task offloading decisions
$\{\xbf_{ij}\}$ and the
communication and CAP processing resource allocation
$\{\rbf_i\}$. This optimization
problem is formulated as follows:
\begin{align}
\min\limits_{\{\xbf_{ij}\},\{\rbf_i\}}
&\quad \sum_{i=1}^N\bigg[\sum_{j=1}^M(E^l_{ij}x^l_{ij}+E^A_{ij}x^a_{ij}+E^C_{ij}x^c_{ij})\nonumber\\
&\quad \quad \ \ +\rho_i\max\{T^{L}_{i}, T^{A}_{i}, T^{C}_{i}\}\bigg]\label{problem_CAP}\allowdisplaybreaks\\
\text{s.t.} &\quad \eqref{a},\eqref{b},\eqref{cap_total},\eqref{problem_CAP.a},\eqref{eq_placement},\nonumber\allowdisplaybreaks\\
&\quad c^u_{i}, c^d_{i}, f^a_{i},\geq 0,\forall i,\label{problem_CAP.b}\allowdisplaybreaks\\
&\quad x^l_{ij}, x^a_{ij}, x^c_{ij}\in \{0,1\},\forall i,
j,\label{problem_CAP.c}
\end{align}
where $E^A_{ij}\triangleq(E^t_{ij}+E^r_{ij}+\alpha C^a_{ij})$ and
$E^C_{ij}\triangleq(E^t_{ij}+E^r_{ij}+\beta C^c_{ij})$ are the
weighted transmission energy and processing costs of offloading and
processing task $j$ of user $i$ to the CAP and cloud, with $\alpha$
and $\beta$ being their relative weights, respectively; also, $T^{L}_{i}$ is the processing delay of tasks processed by the mobile
user $i$ itself,
$T^{A}_{i}$ and $T^{C}_{i}$ are the overall transmission and
remote-processing delays for the tasks of mobile user $i$ processed
at the CAP and cloud, respectively, and $\rho_i$ is the weight on
the task processing delay relative to energy consumption for user $i$.
Comparing with the optimization problem
\eqref{prob_noCAP} in the no-CAP case, the above mixed-integer programming problem
\eqref{problem_CAP} is even more complicated due to the additional
CAP processing cost, $E^A_{ij}$, CAP processing delay, $T^{A}_{i}$,
 and the placement constraint \eqref{eq_placement}.

For optimization problem \eqref{problem_CAP}, we have the
overall local processing delay for each user as
$T^{L}_{i}= \sum_{j=1}^MT^l_{ij}x^l_{ij}$, for all $i$.
However, as similarly discussed in problem \eqref{prob_noCAP}, the overall delay for
CAP processing, $T^{A}_i$, and for cloud processing, $T^{C}_i$, are
not precisely tractable due to multiple offloaded tasks may have overlapping transmission or processing time.
Therefore, we use
both upper and lower bounds of $T^{A}_i$ and $T^{C}_i$ in our
proposed solution and performance benchmarking. We will show later
that, with the proposed MUMTO-C algorithm, the
upper and lower bounds give estimates to the total system cost that
are close to each other.

\subsection{Multi-user Multi-task
Offloading with CAP (MUMTO-C) Algorithm}\label{sec.MUMTO-C}
 To find an efficient solution to the
mixed-integer non-convex programming problem \eqref{problem_CAP}, in
the following, we first propose upper-bound and lower-bound
formulations of both $T^{A}_i$ and $T^{C}_i$, then transform the
optimization problem \eqref{problem_CAP} into a separable QCQP and
the corresponding SDR problem.
Finally, we will propose a three-step MUMTO-C algorithm to obtain the binary offloading
decisions $\{\mathbf{x}_{ij}\}$ and the communication and processing
resource allocation $\{\rbf_i\}$.

\subsubsection{\textbf{Bounds of CAP-Processing and Cloud-Processing Delays}}\label{bound_of_Tc}
Similar to Section \ref{bound_of_Tc_noCAP}, we have the following upper
bounds, i.e., the \textit{worst-case delays}:
\begin{align} \label{worst_Ta_CAP}
T^{{A}_{(U)}}_{i}=\sum_{j=1}^M((T^{t}_{ij}+T^{r}_{ij})(x^a_{ij}+x^c_{ij})+T^{a}_{ij}x^a_{ij}),
\end{align}
\begin{align} \label{worst_Tc_CAP}
\hspace*{-.4em}T^{{C}_{(U)}}_{i}=\sum_{j=1}^M((T^{t}_{ij}+T^{r}_{ij})(x^a_{ij}+x^c_{ij})+(T^{ac}_{ij}+T^{c}_{ij})x^c_{ij}).
\end{align}
In the above expressions, $T^{{A}_{(U)}}_{i}$ and
$T^{{C}_{(U)}}_{i}$ represent the direct summing of the transmission
delays and processing delays without any overlap. They are always
greater than the actual delay given the same offloading decision and
resource allocation.

For performance benchmarking, we will also need the best-case
delays. By separating the offloading delays of all mobile users into
several components and only considering the largest one among them,
the lower bounds of $T^{A}_i$ and $T^{C}_i$ are
\begin{align}\label{best_Ta_CAP}
T^{{A}_{(L)}}_{i}=\max\{T^{u'}_{i}, T^{d'}_{i}, T^{a'}_{i}\},
\end{align}
\begin{align}\label{best_Tc_CAP}
T^{{C}_{(L)}}_{i}=\max\{T^{u}_{i}, T^{d}_{i}, T^{uac}_{i},
T^{dac}_{i}, T^{c'}_{i}\},
\end{align}
where $T^{u'}_{i}= \sum_{j=1}^MT^{t}_{ij}x^a_{ij}$ and $T^{d'}_{i}=
\sum_{j=1}^MT^{r}_{ij}x^a_{ij}$ are the total uplink and downlink
transmission times between the user and the CAP for user $i$'s tasks
processed at the CAP, respectively, $T^{u}_{i}=
\sum_{j=1}^MT^{t}_{ij}x^c_{ij}$ and $T^{d}_{i}=
\sum_{j=1}^MT^{r}_{ij}x^c_{ij}$ are the total uplink and downlink
transmission times between the user and the CAP for user $i$'s tasks
processed at the cloud, respectively,
$T^{uac}_{i}=\sum_{j=1}^MD_{{\textrm{in}}}(ij)x^c_{ij}/r^{ac}$ and
$T^{dac}_{i}= \sum_{j=1}^MD_{{\textrm{out}}}(ij)x^c_{ij}/r^{ac}$ are
the total uplink and downlink transmission times between the CAP and
the cloud for user $i$, respectively, and
$T^{a'}_{i}=\sum_{j=1}^MT^{a}_{ij}x^a_{ij}$ and
$T^{c'}_{i}=\sum_{j=1}^MT^{c}_{ij}x^c_{ij}$ are the total CAP and
cloud processing times for user $i$, respectively.

In the following subsections, we describe the details of the
proposed three-step MUMTO-C algorithm, using the
worst-case delays $T^{{A}_{(U)}}_{i}$ and $T^{{C}_{(U)}}_{i}$ in
optimization problem \eqref{problem_CAP} to obtain an approximate
solution, which gives an upper bound to the actual total system
cost. Furthermore, we show the local optimum property of the
obtained binary offloading decisions $\{\mathbf{x}_{ij}\}$ and
communication and processing resource allocation $\{\rbf_i\}$.
Similarly, $T^{{A}_{(L)}}_{i}$ and $T^{{C}_{(L)}}_{i}$ are used to
obtain a lower bound of the total system cost for performance
benchmarking. Finally, we show in Section \ref{sec_simulation_CAP}
that MUMTO-C achieve actual system cost that is close
to the lower bound of the system cost, and hence is also close to
the optimal system cost.

\subsubsection{\textbf{Step 1: QCQP Transformation and Semidefinite Relaxation}}\label{sec_SDP_CAP}
As mentioned before, optimization problem \eqref{problem_CAP} is more
complicated than problem \eqref{prob_noCAP} due to the availability of the CAP.
In order to obtain the eventual SDR formulation,
we first rewrite the integer
constraint \eqref{problem_CAP.c} as
\begin{align} \label{xy_new_CAP}
x^s_{ij}(x^s_{ij}-1)=0,\quad \forall i, j,
\end{align}
for $s\in\{l,a,c\}$, and replace $T^{A}_{i}$ and $T^{C}_{i}$ in
\eqref{problem_CAP} with $T^{{A}_{(U)}}_{i}$ and
$T^{{C}_{(U)}}_{i}$, respectively.
Following the similar procedure
in Section \ref{sec_QCQP_noCAP}, we move the delay term from the objective to
the constraints by using additional auxiliary variables
$t_i$, and rewrite \eqref{problem_CAP} as
\begin{align}
\hspace*{-.25cm}\min\limits_{\{\mathbf{x}_{ij}\},\{\rbf_i,t_i\}}
& \sum_{i=1}^N\bigg[\sum_{j=1}^M(E^l_{ij}x^l_{ij}+E^A_{ij}x^a_{ij}+E^C_{ij}x^c_{ij})+\rho_it_i\bigg]\label{problem_CAP_new}\allowdisplaybreaks\\
\text{s.t.} \quad & \sum_{j=1}^MT^l_{ij}x^l_{ij}\leq t_i,\ \forall i\nonumber,\allowdisplaybreaks\\
&
\sum_{j=1}^M\left(\frac{D_{{\textrm{in}}}(ij)}{\eta^u_ic^u_i}+\frac{D_{{\textrm{out}}}(ij)}{\eta^d_ic^d_i}\right)(x^a_{ij}+x^c_{ij})\nonumber\\
&+\sum_{j=1}^M\frac{Y(ij)}{f^a_i}x^a_{ij}\leq t_i,\ \forall i\nonumber,\\
&
\sum_{j=1}^M\left(\frac{D_{{\textrm{in}}}(ij)}{\eta^u_ic^u_i}+\frac{D_{{\textrm{out}}}(ij)}{\eta^d_ic^d_i}\right)(x^a_{ij}+x^c_{ij})\nonumber\\
&+\sum_{j=1}^M(T^{ac}_{ij}+T^{c}_{ij})x^c_{ij}\leq t_i,\ \forall i\nonumber,\allowdisplaybreaks\\
&
\eqref{a},\eqref{b},\eqref{cap_total},\eqref{problem_CAP.a},\eqref{eq_placement},\eqref{problem_CAP.b},
\text{~and} \ \eqref{xy_new_CAP}. \nonumber
\end{align}

Comparing   problem \eqref{problem_CAP_new} with
problem \eqref{prob_noCAP_new}, we observe that they share a similar
structure. Therefore, we can apply a similar procedure to
transform problem \eqref{problem_CAP_new} into a non-convex separable
QCQP problem that is similar to problem \eqref{Eq_homo_QCQP}, with the optimization vector now defined by  $\vbf_i\triangleq [\tilde{\wbf}_i^T,1]^T$, where  $\tilde{\wbf}_i\triangleq[\xbf_{i1}^T,\cdots,\xbf_{iM}^T,c_i^u, D_i^u, c_i^d,D_i^d,f_i^a, D_i^a,t_i]^T$, with  $D_i^u$, $D_i^d$ and $D_i^a$ being the auxiliary variables introduced corresponding to the uplink transmission time, downlink transmission time, and the CAP\ processing time, respectively. Auxiliary variables  $D_i^u$ and $D_i^d$ are similarly defined as in \eqref{constraint_D^u} and \eqref{constraint_D^d}, except that $x_{ij}$  in \eqref{constraint_D^u} and \eqref{constraint_D^d} is now replaced by $(x^a_{ij}+x^c_{ij})$. Similar to these two constraints, the new auxiliary variable $D_i^a$ for the CAP processing time also introduces a new constraint   $\sum_{j=1}^M{Y(ij)}x^a_{ij}/{f^a_i}\le D_i^a$. Using the separable SDR approach, we solve the relaxed separable SDP problem that is similar to problem \eqref{Eq_SDP_QCQP_noCAP}, with optimization matrix defined by $\Vbf_i=\vbf_i\vbf_i^T$ with size $(3M+8)\times (3M+8)$. The details are omitted to avoid redundancy.

Denote  $\{\Vbf_i^*\}$ as the optimal
solution of the corresponding separable SDR problem for the
optimization problem \eqref{problem_CAP_new}. We need to recover a
rank-one solution from  $\{\Vbf_i^*\}$ for problem
\eqref{problem_CAP_new}. However, the reconstruction of binary offloading decision $\{\xbf_{ij}\}$ in Section \ref{sec.MDBSDR}, as part of the MUMTO algorithm, cannot be directly applied to
find a feasible solution for problem \eqref{problem_CAP_new} due to
the additional placement constraint \eqref{eq_placement} for each user's tasks. To deal with this challenge, in the following, we
propose an modified method, termed \textit{ MUMTO SDR with CAP (SDR-C)}, to obtain the binary offloading decisions
$\{\mathbf{x}_{ij}\}$ and the corresponding optimal communication
resource allocation $\{\rbf_i\}$ from $\{\Vbf_i^*\}$.

Define $\xbf \triangleq [\xbf_1^T,\ldots, \xbf_N^T]^T$, where
$\xbf_i \triangleq [\mathbf{x}_{i1},\ldots, \mathbf{x}_{iM}]^T$, for
all $i$. As similarly discussed in Section~\ref{sec.MDBSDR}, $\Vbf_i(3M+8,k)=\vbf_i(k)$, for $k=1,\cdots,3M$, which correspond to\ offloading decision $\xbf_i$ for user $i$.   It can be proven that optimal solution $\Vbf^*_i(3M+8,k)\in [0,1]$, for $k=1,...,3M$. Denote  $\mathbf{p}_{ij} \triangleq [p^l_{ij},
p^a_{ij}, p^c_{ij}]^T$ and $\pbf_i\triangleq[\mathbf{p}_{i1}^T,\ldots,
\mathbf{p}_{iM}^T]^T \triangleq[\Vbf^*_i(3M+8,1),\cdots,\Vbf^*_i(3M+8,3M)]^T$. Then, we have each element in $\mathbf{p}_i$ having its  value within $ [0,1]$, $\forall i$. We recover the feasible
decisions $\xbf_i^{\text{sdr}}$ using $\pbf_i$ as follows: for $j=1,\cdots,M$, set
\begin{equation}\label{eq_sdr_recovery}
\hspace*{-.15cm}\mathbf{x}_{ij}^{\text{sdr}}
\hspace*{-.075cm}=\hspace*{-.1cm}
\begin{cases}
\hspace*{-.05cm}[1,0,0]^T\hspace*{-.05cm},&\hspace*{-.25cm}\text{if}\ \hspace*{-.05cm}  \smash{\displaystyle\max_{s\in \{l,a,c\}}}  p_{ij}^s\hspace*{-.05cm}=\hspace*{-.05cm}p^l_{ij}\ \hspace*{-.05cm}\text{(local processing)}\hspace*{-.05cm}\\
\hspace*{-.05cm}[0,1,0]^T\hspace*{-.05cm},&\hspace*{-.25cm}\text{if}\ \hspace*{-.05cm} \smash{\displaystyle\max_{s\in \{l,a,c\}}}  p_{ij}^s\hspace*{-.05cm}=\hspace*{-.05cm}p^a_{ij}\ \hspace*{-.05cm}\text{(CAP processing)}\hspace*{-.05cm}\\
\hspace*{-.05cm}[0,0,1]^T\hspace*{-.05cm},&\hspace*{-.25cm}\text{if}\ \hspace*{-.05cm} \smash{\displaystyle\max_{s\in \{l,a,c\}}}  p_{ij}^s\hspace*{-.05cm}=\hspace*{-.05cm}p^c_{ij}\ \hspace*{-.05cm}\text{(cloud processing)},\hspace*{-.05cm}\\
\end{cases}
\end{equation}
The overall offloading decision is obtained as
$\xbf^{\text{sdr}}=[(\xbf_1^{\text{sdr}})^T,\ldots,
(\xbf_N^{\text{sdr}})^T]^T$.

After obtaining the offloading decision $\xbf^{\text{sdr}}$,
optimization problem \eqref{problem_CAP} is reduced to the optimization of
computation and communication resource allocation $\{\rbf_i\}$,
which is given by
\begin{align}
\min\limits_{\{\rbf_i\}}
&\quad \bigg(E+\sum_{i=1}^N\rho_i\max\{T^{L}_{i}, T^{{A}_{(U)}}_{i}, T^{{C}_{(U)}}_{i}\}\bigg)\label{h}\allowdisplaybreaks\\
\text{s.t.} &\quad \eqref{a},\eqref{b},\eqref{cap_total},\eqref{problem_CAP.a}, \text{~and} \ \eqref{problem_CAP.b},\nonumber
\end{align}
where
$E\triangleq\sum_{i=1}^N\sum_{j=1}^M(E^l_{ij}x^l_{ij}+E^A_{ij}x^a_{ij}+E^C_{ij}x^c_{ij})$
is a constant value once $\{\mathbf{x}_{ij}\}$ are given. Similar to problem \eqref{prob_resource_noCAP}, problem \eqref{h} is convex, so it can be solved
optimally.

\subsubsection{\textbf{Step 2: Improvement to SDR-C by Alternating Optimization (AO)}}
After obtaining a feasible solution
$\{\xbf^{\text{sdr}},\{\rbf^{\text{sdr}^*}_i\}\}$ from the
SDR-C step above, to further reduce the overall system
cost, in the following we introduce an iterative alternating
optimization method to further improve the offloading decision, by
using $\{\xbf^{\text{sdr}},\{\rbf^{\text{sdr}^*}_i\}\}$ as the
starting point of iteration.

As mentioned above, given any offloading decision, the optimization
problem \eqref{problem_CAP} is reduced to the resource allocation problem
\eqref{h}, which is convex and the optimal resource allocation can
be obtained. On the other hand, once the resource allocation
$\{\rbf_i\}$ is given, the optimization problem \eqref{problem_CAP} is reduced
to the optimization of offloading decisions $\{\mathbf{x}_{ij}\}$ as
follows:
\begin{align}\label{offloading_decision_only}
\min\limits_{\{\mathbf{x}_{ij}\}}
&\quad \sum_{i=1}^N\bigg[\sum_{j=1}^M(E^l_{ij}x^l_{ij}+E^A_{ij}x^a_{ij}+E^C_{ij}x^c_{ij})\nonumber\\
&\quad \quad \ \ +\rho_i\max\{T^{L}_{i}, T^{{A}_{(U)}}_{i}, T^{{C}_{(U)}}_{i}\}\bigg]\allowdisplaybreaks\\
\text{s.t.} &\quad \eqref{eq_placement} \text{~and} \
\eqref{xy_new_CAP}.\nonumber\allowdisplaybreaks
\end{align}
The offloading decision problem \eqref{offloading_decision_only} is
an integer programming problem. However, it can be separated into
$N$ independent sub-problems, where each sub-problem only considers
the offloading decision of one user. As shown in
\cite{chen2015spawc}, this can be solved near-optimally by either
using an SDR approach or relaxing the integer constraints to
interval constraints.
Since the optimization problem \eqref{problem_CAP} can be separated into two
sub-problems \eqref{h} and \eqref{offloading_decision_only}. We
propose the following alternating optimization procedure to further
reduce the total system cost.

Set
$(\mathbf{x}^{\text{ao}^*},\{\rbf^{\text{ao}^*}_i\})=(\xbf^{\text{sdr}},\{\rbf^{\text{sdr}^*}_i\})$
as the initial point. At each iteration:
 \begin{enumerate}
 \item[i)] Solve problem \eqref{offloading_decision_only} based on
$\{\rbf^{\text{ao}^*}_i\}$ to find the corresponding offloading
decision $\xbf^{\text{ao}'}$.
\item[ii)] Solve problem \eqref{h} based on $\xbf^{\text{ao}'}$ to
find the minimum system cost and the corresponding resource
allocation $\{\rbf^{\text{ao}'}_i\}$. If this provides a lower total
system cost, update
$(\mathbf{x}^{\text{ao}^*},\{\rbf^{\text{ao}^*}_i\})=(\mathbf{x}^{\text{ao}'},\{\rbf^{\text{ao}'}_i\})$.
 \end{enumerate}
Repeat steps i and ii until the the total system cost cannot be
further decreased. Then output the solution of the alternating
optimization procedure as
$(\xbf^{\text{ao}^*},\{\rbf^{\text{ao}^*}_i\})$.

Note that, despite the approximation in solving
\eqref{offloading_decision_only}, since we only accept a better
solution in each iteration, and the system cost is lower bounded, AO
always converges. Furthermore, by design, the solution
$(\xbf^{\text{ao}^*},\{\rbf^{\text{ao}^*}_i\})$ is better than or at
least as good as $(\xbf^{\text{sdr}},\{\rbf^{\text{sdr}^*}_i\})$.

\subsubsection{\textbf{Step 3: Sequential Tuning (ST) to Reach Local Optimum}}
 In this step, we propose an iterative procedure
starting from $\{\xbf^{\text{ao}^*},\{\rbf^{\text{ao}^*}_i\}\}$,
termed sequential tuning, to further reduce the system cost and
eventually achieve a local optimum for \eqref{problem_CAP}.

Set
$(\mathbf{x}^{\text{st}^*},\{\rbf^{\text{st}^*}_i\})=(\xbf^{\text{ao}^*},\{\rbf^{\text{ao}^*}_i\})$
as the initial point. At each iteration:
 \begin{enumerate}
 \item[i)] Randomly order the lists of all users and their tasks.
 \item[ii)] Go through the user list one by one. For each examined user, sequentially check each
 of its tasks for the three possible offloading decisions, while the offloading decisions
 of all other tasks of all users remain unchanged. For each offloading decision, find the total system
 cost by solving problem \eqref{h}. As soon as some user $i$ is found to admit a lower total system cost
 by changing the offloading decision of one of its tasks, update $(\mathbf{x}^{\text{st}^*},\{\rbf^{\text{st}^*}_i\})$ to the new offloading decision
 and resource allocation that give the lower cost, and exit the iteration.
 \end{enumerate}
Repeat steps i and ii until $\mathbf{x}^{\text{st}^*}$ converges,
i.e., no change
 for $\mathbf{x}^{\text{st}^*}$ can be made. Then output the solution of the sequential turning procedure as
$(\mathbf{x}^{\text{st}^*},\{\mathbf{r}_i^{\text{st}^*}\})$.

The above procedure is guaranteed to converge. This is because there
is a finite
 number of possible values for $\mathbf{x}^{\text{st}}_i$. The iteration eventually will reach some
$(\mathbf{x}^{\text{st}^*},\{\mathbf{r}_i^{\text{st}^*}\})$, where
the total system cost cannot be further reduced by modifying any
user's offloading decision (and corresponding resource allocation).
It is straightforward to show that
$(\mathbf{x}^{\text{st}^*},\{\mathbf{r}_i^{\text{st}^*}\})$ is a
local optimum of problem \eqref{problem_CAP}, since it gives the
lowest system cost in the joint binary-valued neighborhood of
$\mathbf{x}$ and neighborhood of $\{\mathbf{r}_i\}$. This result is
stated in the following proposition.
\begin{proposition}\label{thm.ST}
The solution $(\mathbf{x}^{\text{st}^*},\{\mathbf{r}_i^{\text{st}^*}\})$ obtained
from the sequential tuning procedure is a locally optimal solution
to the original non-convex optimization problem \eqref{problem_CAP}.
\end{proposition}
\subsubsection{\textbf{Overall MUMTO-C Algorithm}}

We summarize the above three-step MUMTO-C algorithm
in Algorithm \ref{algorithm}.

\begin{algorithm}[!t]
\small
  \caption{MUMTO-C Algorithm}
  \begin{algorithmic}[1]
  \Statex \textbf{Step 1: Initial offloading solution via SDR-C}
  \State Transform the original problem \eqref{problem_CAP} into the SDR
  problem and obtain the optimal solution $\{\Vbf_i^*\}$.
    \State Extract $\Vbf^*_i(3M+8,k)$, for $k=1,...,3M$, from $\mathbf{V}_i^*$.
 \State Record the values of $\Vbf^*_i(3M+8,k)$, for $k=1,...,3M$, by $\pbf_i=[\mathbf{p}_{i1}^T,\ldots,
\mathbf{p}_{iM}^T]^T$, where $\mathbf{p}_{ij} = [p^l_{ij}, p^a_{ij},
p^c_{ij}]^T$.
      \State Set $\xbf^{\text{sdr}}=[(\xbf^{\text{sdr}}_1)^T,\ldots,
(\xbf^{\text{sdr}}_N)^T]^T$, where $\mathbf{x}^{\text{sdr}}_{i}$ is
given by \eqref{eq_sdr_recovery}, and solve problem \eqref{h} based
on $\xbf^{\text{sdr}}$.
    \Statex \textbf{Step 2: Alternating optimization (AO)}
    \State Set $(\mathbf{x}^{\text{ao}^*},\{\rbf^{\text{ao}^*}_i\})=(\xbf^{\text{sdr}},\{\rbf^{\text{sdr}^*}_i\})$, and record the corresponding total system cost as
$J^{\text{ao}^*}$; set $\mathrm{AO}=\mathrm{False}$.
    \While{$\mathrm{AO}==\mathrm{False}$}
   \State Solve problem \eqref{offloading_decision_only} based on
$\{\rbf^{\text{ao}^*}_i\}$ to find the corresponding
\Statex $\ \ \ \ $offloading
decision $\xbf^{\text{ao}'}$;
 \State Solve problem \eqref{h} based on $\xbf^{\text{ao}'}$ to
find the minimum system
\Statex $\ \ \ \ $cost $J^{\text{ao}'}$ and
$\{\mathbf{r}_i^{\text{ao}'}\}$;
    \If{$J^{\text{ao}'}< J^{\text{ao}^*}$}
    \State Set $(\mathbf{x}^{\text{ao}^*},\{\rbf^{\text{ao}^*}_i\})=(\mathbf{x}^{\text{ao}'},\{\rbf^{\text{ao}'}_i\})$,
   $J^{\text{ao}^*}=J^{\text{ao}'}$;
    \Else
    \State Set $\mathrm{AO}=\mathrm{True}$; \Comment{Exit while loop}
    \EndIf
    \EndWhile
\Statex \textbf{Step 3: Sequential tuning (ST)}
    \State Set $(\mathbf{x}^{\text{st}^*},\{\rbf^{\text{st}^*}_i\})=(\xbf^{\text{ao}^*},\{\rbf^{\text{ao}^*}_i\})$,
    and record the corresponding total system cost as
$J^{\text{st}^*}$; set $\mathrm{ST}=\mathrm{False}$.
\While{$\mathrm{ST}==\mathrm{False}$}
    \State Randomly order the lists of all users and their tasks;
set user
\Statex $\ \ \ \ $ index $n=1$; set task index $m=1$;
    \While{$n\leq N$ and $m\leq M$}
    \State While keeping $\mathbf{x}^{\text{st}^*}_{n'm'}$ unchanged  for all $(n', m')$ except
     \Statex $\ \ \ \ \ \ \ \ $ $(n', m')=(n, m)$,
    inspect the three possible offloading
    \Statex $\ \ \ \ \ \ \ \ $ choices of
    $\mathbf{x}^{\text{st}^*}_{nm}$; find their respective total
    system costs
    \Statex $\ \ \ \ \ \ \ \ $ by solving problem \eqref{h}; set $J^{\mathrm{st}'}$ as the minimum cost
    \Statex $\ \ \ \ \ \ \ \ $  among these three choices, and record the
     corresponding
     \Statex $\ \ \ \ \ \ \ \ $ solution as $(\mathbf{x}^{\mathrm{st}'}, \{\mathbf{r}_i^{\mathrm{st}'}\})$;

     \If{$J^{\text{st}'}< J^{\text{st}^*}$}
    \State Set
    $(\mathbf{x}^{\text{st}^*},\{\rbf^{\text{st}^*}_i\})=(\mathbf{x}^{\text{st}'},\{\rbf^{\text{st}'}_i\})$, $J^{\text{st}^*}=J^{\text{st}'}$;
    \Statex $\ \ \ \ \ \ \ \ \ \ \ \ \ $$n \leftarrow N+1$;
    \ElsIf{$n==N$ and $m==M$}
    \State $n \leftarrow N+1$; $\mathrm{ST}=\mathrm{True}$;
    \Comment{No change of $\xbf^{\text{st}^*}$ can \Statex $\ \ \ \ \ \ \ \ \ \ \ \ \ \ \ \ \ \ \ \ \ \ \ \ \ \ \ \ \ \ \ \ \ \ \ \ \ \ \ \ \ \ \ \ \ \ \ \ $ be found; exit}
    \ElsIf{$n<N$ and $m==M$}
    \State $n \leftarrow n+1$; $m \leftarrow 1$;
    \Else
    \State $m \leftarrow m+1$;
    \EndIf
    \EndWhile
    \EndWhile

    \State Output: The offloading decision $\mathbf{x}^{\text{st}^*}$ and the corresponding resource allocation $\{\mathbf{r}_i^{\text{st}^*}\}$.
  \end{algorithmic}
  \label{algorithm}

\end{algorithm}

Even though each of the SDR-C, AO, and ST steps above can be used
separately to provide a feasible solution to the original
optimization problem \eqref{problem_CAP}, when combined together in
the proposed manner, they each serves an important role in the
overall algorithm design. First, SDR-C provides a suitable starting
point for AO. Without it, i.e., if we start the AO iteration from
some randomly chosen point in the solution space, as shown in
Section \ref{sec_simulation_CAP}, AO can converge to some highly
sub-optimal solution. Second, with an appropriate starting point, AO
pushes the solution to one that is closer to the optimum. This
provide a suitable starting point for ST, which helps reduce the
number of iterations in ST. This is an important step, since each of
the ST iterations can be computationally expensive, as it
potentially may require searching over a large number of tasks.
Finally, ST further improves the solution, and more importantly, it
guarantees that the final solution is a local optimum.
Further
numerical evaluation of the roles and contributions of each of these
steps is given in Section \ref{sec_simulation_CAP}.

\subsection{Lower Bound on the Optimal Solution}
Similar to the case of MUMTO in Section \ref{sec_lowerbound_noCAP},
to study the performance of
MUMTO-C compared with an optimal solution, we find a
lower bound of the optimal solution by introducing a new
optimization problem in which $T^{{A}_{(L)}}_i$ and
$T^{{C}_{(L)}}_i$ are used instead as

\begin{align}\label{Eq_new_objective_CAP}
\hspace*{-1em}\min\limits_{\{\mathbf{x}_{ij}\},\{\rbf_i\}}
&\quad \sum_{i=1}^N\bigg[\sum_{j=1}^M(E^l_{ij}x^l_{ij}+E^A_{ij}x^a_{ij}+E^C_{ij}x^c_{ij})\nonumber\\
&\quad \quad \ \ +\rho_i\max\{T^{L}_{i},T^{{A}_{(L)}}_i,T^{{C}_{(L)}}_i\}\bigg]\\
\text{s.t.}
&\quad \eqref{a},\eqref{b},\eqref{cap_total},\eqref{problem_CAP.a},\eqref{eq_placement},\eqref{problem_CAP.b}, \text{~and} \ \eqref{xy_new_CAP}.\nonumber
\end{align}
Under the same offloading decisions and resource
allocation, the objective function in \eqref{Eq_new_objective_CAP} is
always lower than the actual cost.
We apply the same approach to solve the corresponding separable SDR problem of the above non-convex problem \eqref{Eq_new_objective_CAP}.
Since the optimal objective of this SDR problem is
smaller than the optimal objective of \eqref{Eq_new_objective_CAP}, it can serve as a lower bound to the minimum total system
cost defined by the original optimization problem \eqref{problem_CAP}.

In Section \ref{sec_simulation}, we show that the proposed
MUMTO-C method provides not only a local optimum
solution but also nearly optimal performance compared with the lower
bound.

\section{Performance Evaluation}\label{sec_simulation}
In this section, we provide computer simulation to
study the performance of both proposed MUMTO and MUMTO-C offloading
solutions, respectively, under different parameter settings.

\subsection{Simulation Setup}

The following default parameter values are used unless specified
otherwise later. We adopt the mobile device characteristics from
\cite{miettinen2010}, which is based on a Nokia smart device. According to Tables 1 and 3 in
\cite{miettinen2010}, the mobile device has CPU rate $500\times
10^6$ cycles/s and unit processing energy consumption
$\frac{1}{730\times 10^6}$ J/cycle. The local computation time per
bit is $4.75\times 10^{-7}$ s and local processing energy
consumption per bit is $3.25\times 10^{-7}$ J. We consider the x264
CBR encode application, which requires 1900 cycles/byte
\cite{miettinen2010}, i.e., $Y(ij)=1900D_\mathrm{in}(ij)$. The input
and output data sizes of each task are assumed to be uniformly
distributed from $10$ to $30$MB and from $1$ to $3$MB, respectively.

The total transmission bandwidth between the mobile users and the
CAP is set to $40$ MHz, with no additional limit on the uplink or
downlink, and the transmission and receiving energy consumptions of
the mobile user are both $1.42\times 10^{-7}$ J/bit as indicated in
Table 2 in \cite{miettinen2010}. For simplicity, we set
$\eta^u_{i}=\eta^d_{i}=3.5\ \mathrm{b}/\mathrm{s}/\mathrm{Hz}$ for
all $i$.  When tasks are sent from the CAP to
the cloud, the transmission rate $r^{ac}$ is $15$ Mpbs. The cloud
and CAP usage costs are assumed to be
$C^c_{ij}=D_{\textrm{in}}(ij)+\lambda_1/f^{c}+\lambda_2/C_{\textrm{UL}}+\lambda_3/C_{\textrm{DL}}$
and
$C^a_{ij}=D_{\textrm{in}}(ij)+\lambda_1/f_{A}+\lambda_2/C_{\textrm{UL}}+\lambda_3/C_{\textrm{DL}}$,
respectively, where $\lambda_1=10^{18}$ $\mathrm{bit}\times
\mathrm{cycle/s}$ and $\lambda_2=\lambda_3=10^{16}$
$\mathrm{bit}\times \mathrm{MHz}$, which accounts for the input data
size, processing rate, and uplink and downlink capacities.

The default values for other parameters are summarized in Table
\ref{table_simulation}. Unless specified otherwise, these default
values are used in the figures below. All simulation results are
obtained by averaging over 100 realizations of the input and output
data sizes of each task.

\begin{table}[t]
\caption{Default parameter settings.} \vspace*{-1em}
\centering
\small
\begin{tabular}{l| l}
\hline\hline 
\textbf{Description}& \textbf{Default Value} \\
\hline 
number of users $N$ & 5\\
number of tasks per user $M$ & 4\\
total CAP processing rate $f_A$ &  $10\times 10^9\ \mathrm{cycle/s}$\\
cloud processing rate $f^{c}$ & $10\times 10^9\ \mathrm{cycle/s}$\\
weight on CAP usage cost $\alpha$ & $1.5\times10^{-7}$ J/bit\\
weight on cloud usage cost $\beta$ & $2.5\times10^{-7}$ J/bit\\
weight on delays $\rho_i$ & $1$ J/s\\
 \hline\hline
\end{tabular}
\label{table_simulation}
\end{table}

\begin{figure}[t!]
\centering
\includegraphics [scale =0.42]{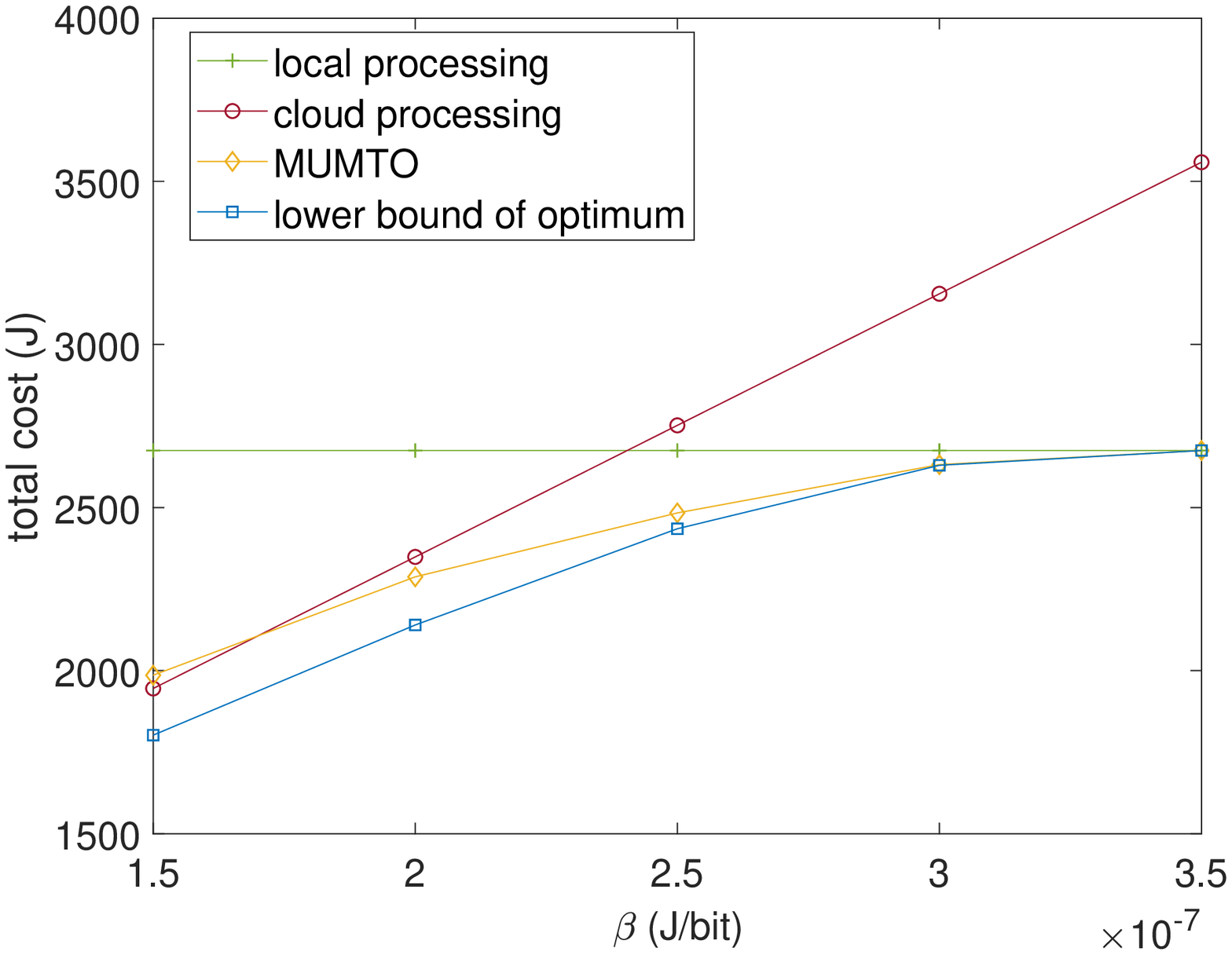}\vspace{-0.25cm}
\caption{ Total system cost versus $\beta$
 without CAP.} \label{fig:beta}
\includegraphics [scale =0.42]{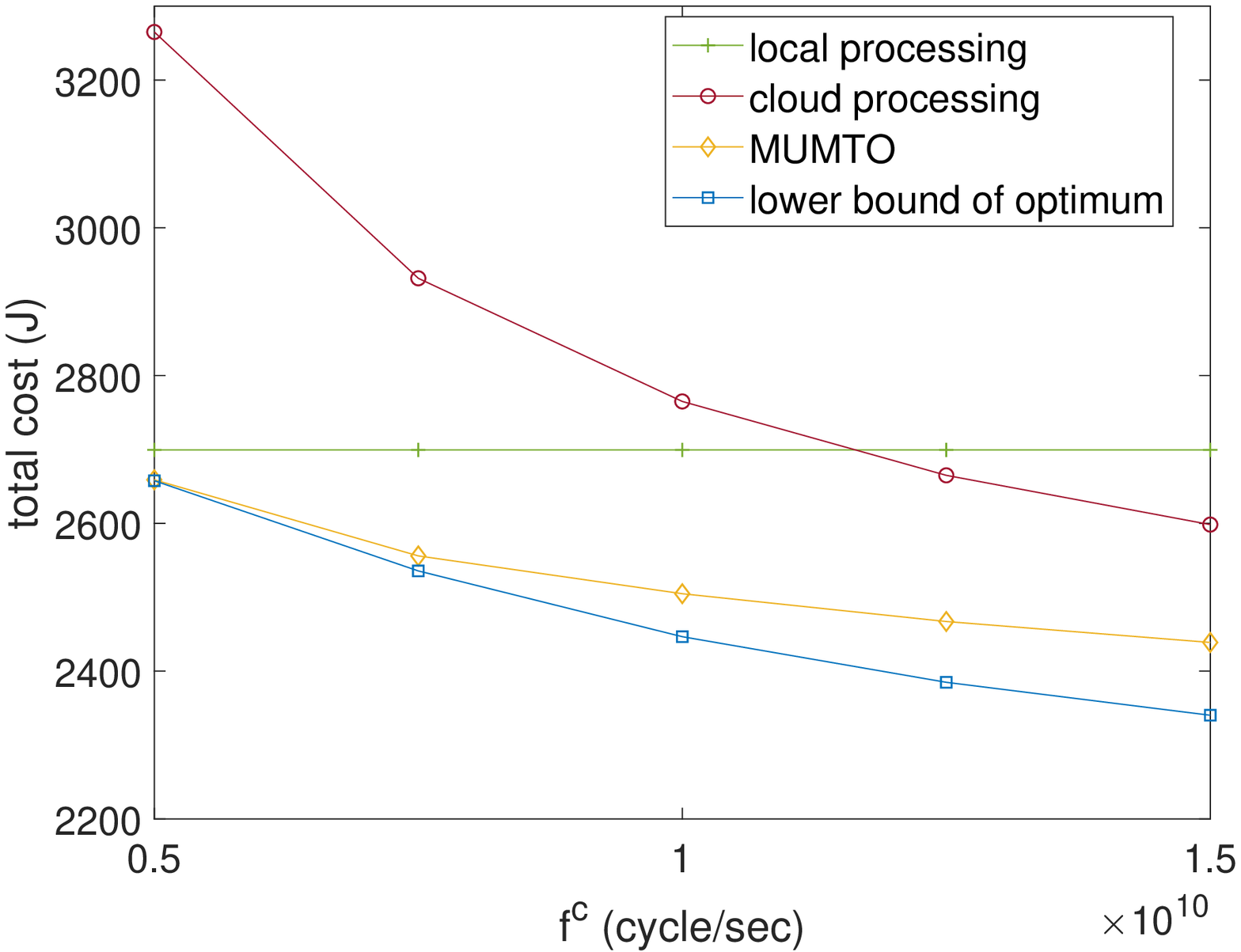}\vspace{-0.25cm}
 \caption{ Total system cost versus cloud CPU rate
$f^c$ without CAP.} \label{fig:f_C}
\end{figure}

\begin{figure}[t!]
    \centering
\includegraphics [scale =0.42]{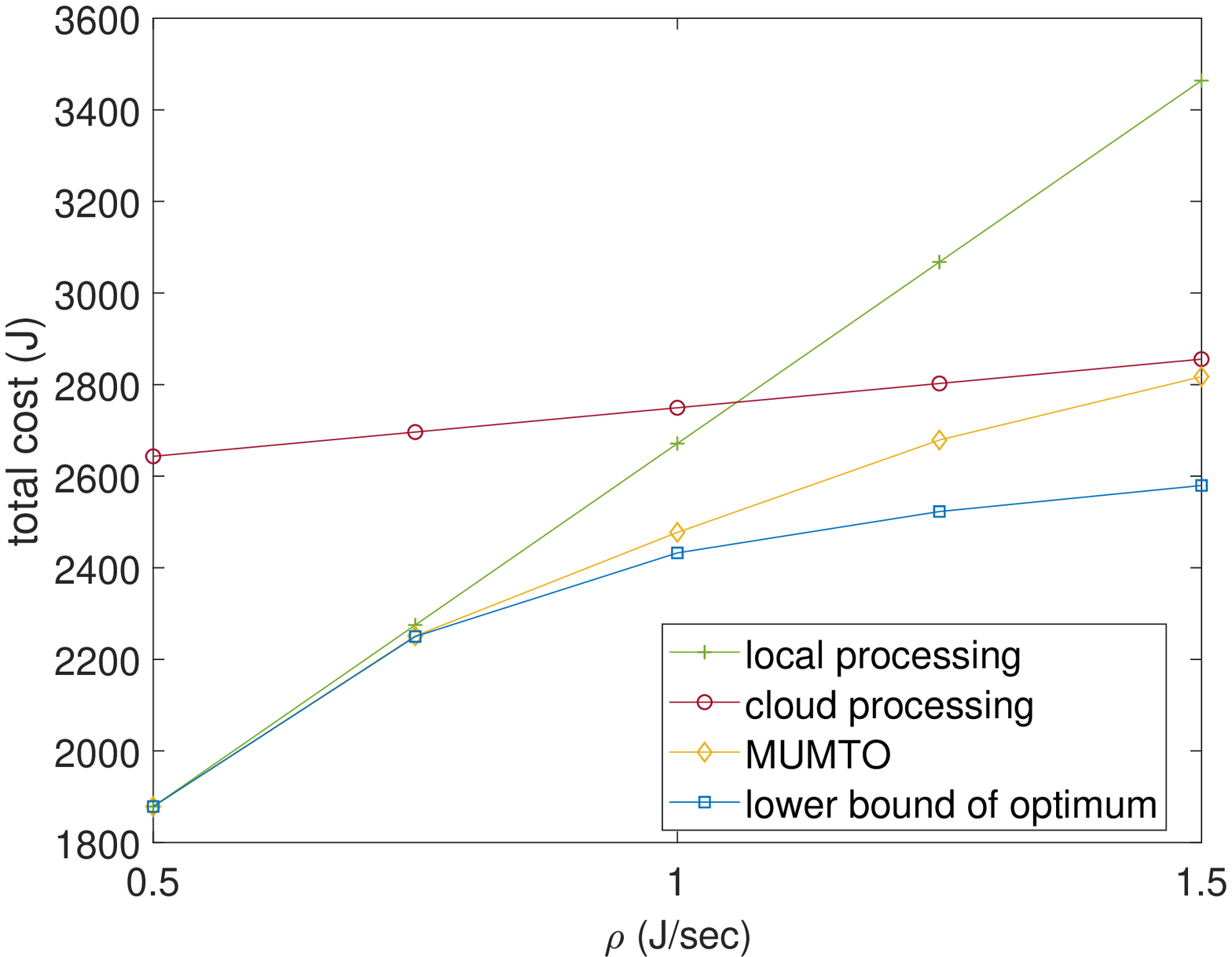}\vspace{-0.25cm}
\caption{ Total system cost versus $\rho$ ($\rho_i$)
 without CAP.} \label{fig:rho}
\includegraphics [scale =0.42]{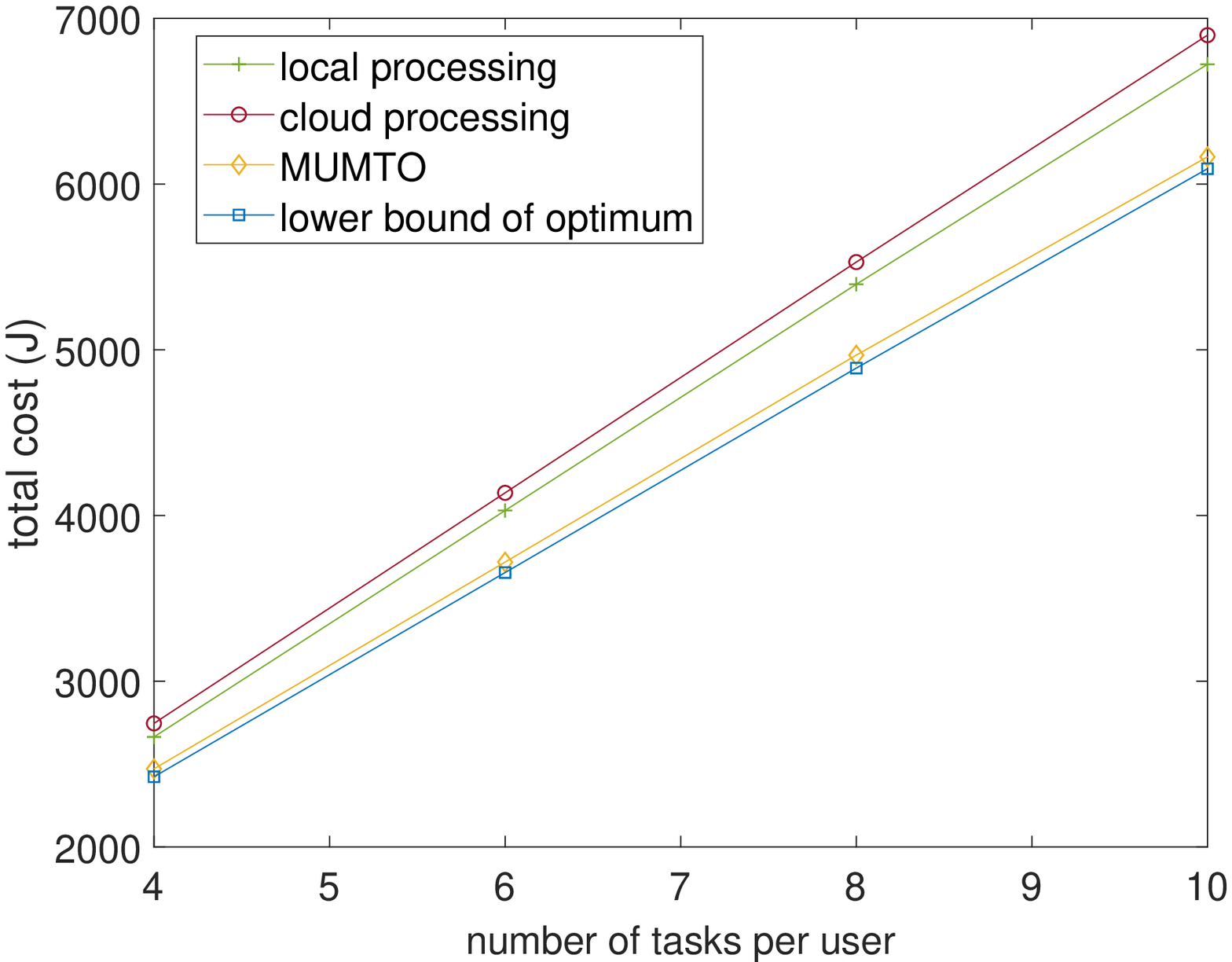}\vspace{-0.25cm}
\caption{ Total system cost versus the number of tasks $M$ per user without
CAP.} \label{fig:task}
\end{figure}

\subsection{Performance Evaluation for MUMTO without CAP}\label{sec_simulation_noCAP}

For comparison, we also consider the following methods: 1) the
\textit{local processing only} scheme where all tasks are processed
by mobile users, 2) the \textit{cloud processing only} scheme where
all tasks are offloaded to the cloud and the cost is obtained based
on $T^{C_{(L)}}_i$, 3) the \textit{lower bound of optimum}, which is
obtained from the optimal objective value of the SDR of problem
\eqref{Eq_new_objective_noCAP}. Notice that in all figures the real
cost under the same offloading decision and resource allocation will
always fall between the costs of the proposed MUMTO and the lower
bound of optimum.

In Fig.~\ref{fig:beta}, we show the system cost vs. the weight
$\beta$ on the system utility cost. When $\beta$ becomes large, all
tasks are more likely to be processed by mobile users themselves.
Both MUMTO and the lower bound of optimum in this case
converge to the local processing only method. Though the existence
of the cloud can provide additional computation capacity, the
processing time at the cloud depends on the cloud CPU rate $f^c$
assigned to each user. In Fig.~\ref{fig:f_C}, we plot the total
system cost vs.~$f^c$. As expected, a more powerful per-user cloud
CPU can dramatically increase system performance, and
MUMTO converges to the local processing only method when
the per-user cloud CPU rate is too slow to help.

In Fig.~\ref{fig:rho}, we study the system cost under various values
of weight $\rho_i=\rho$ on the delays. We observe that MUMTO
substantially outperforms all other methods. Finally, we examine the
scalability of MUMTO. Fig.~\ref{fig:task} plot the total system cost
vs. the number of tasks $M$ per user. We see that MUMTO is close to
the lower bound of optimum, indicating that it is nearly optimal for
all $M$ values.

\subsection{Performance Evaluation for MUMTO-C with CAP}\label{sec_simulation_CAP}

\subsubsection{\textbf{Contribution of the Algorithm Components}}

\begin{figure}[t!]
    \centering
\includegraphics [scale =0.42]{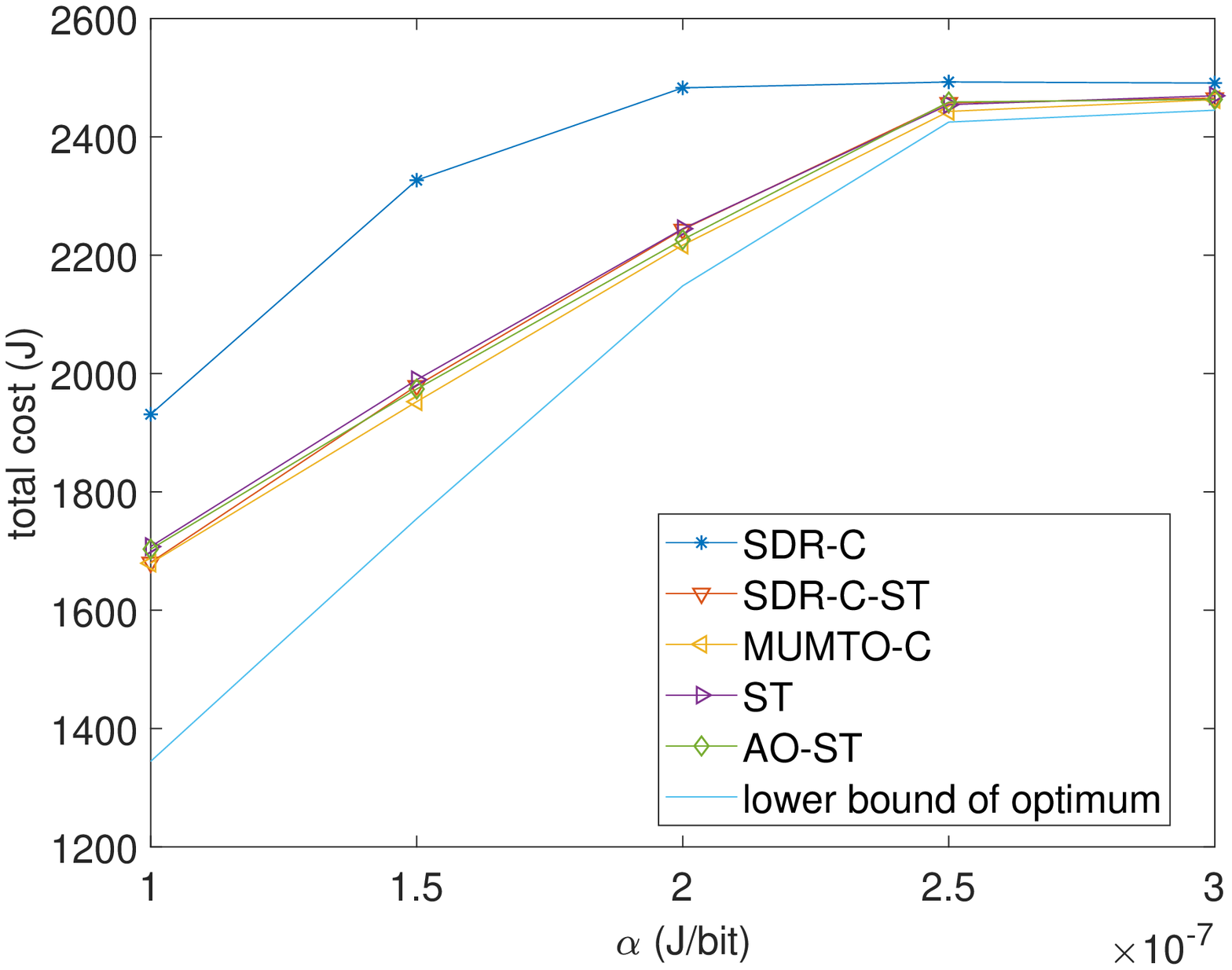}
\caption{Total system cost versus $\alpha$ with CAP.}
\label{fig:alpha_SDR_CAP}
\includegraphics [scale =0.42]{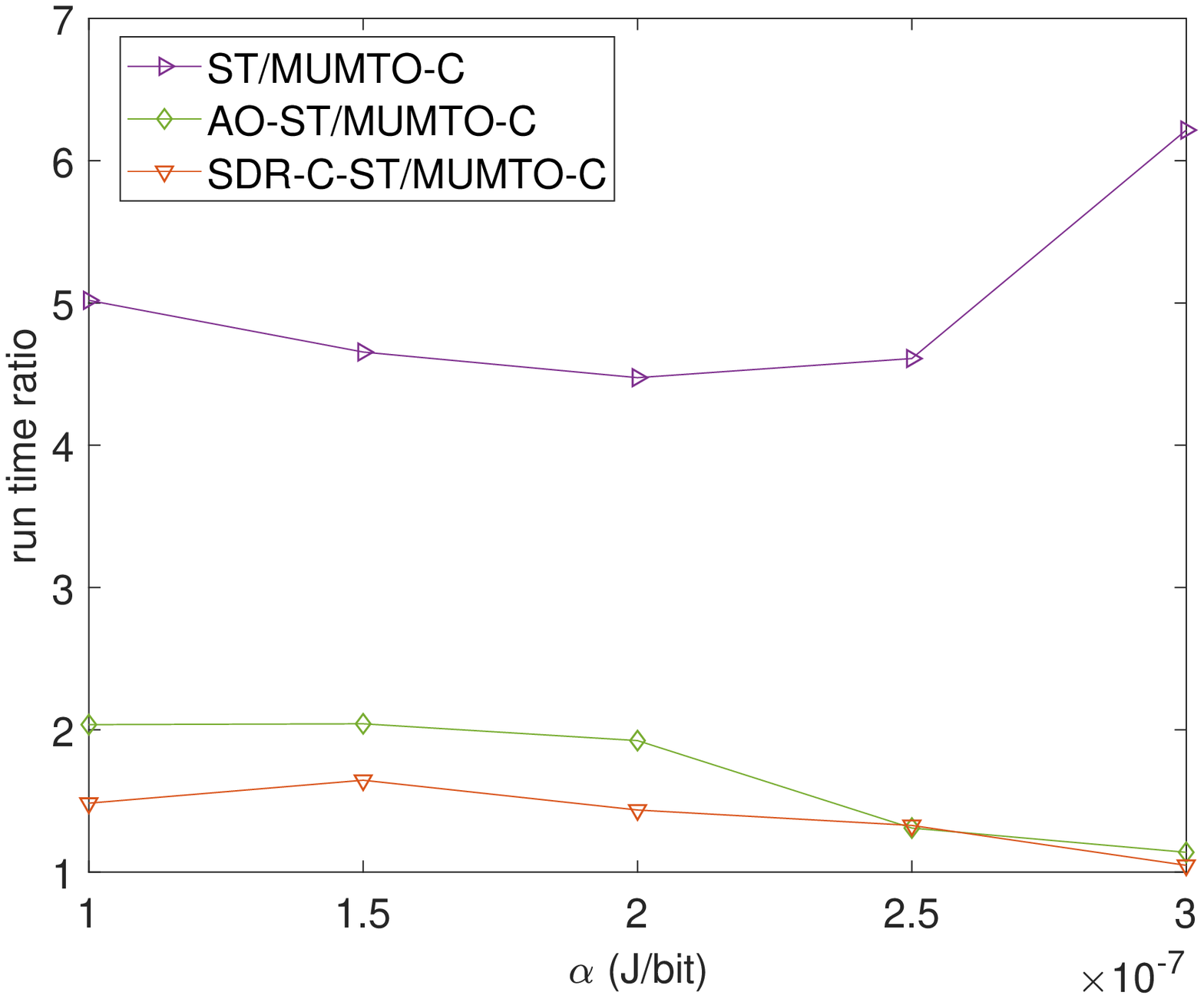}
\caption{Run time ratio versus $\alpha$ with CAP.}
\label{fig:alpha_time_SDR_CAP}
\end{figure}
\begin{figure}[t!]
    \centering
\includegraphics [scale =0.42]{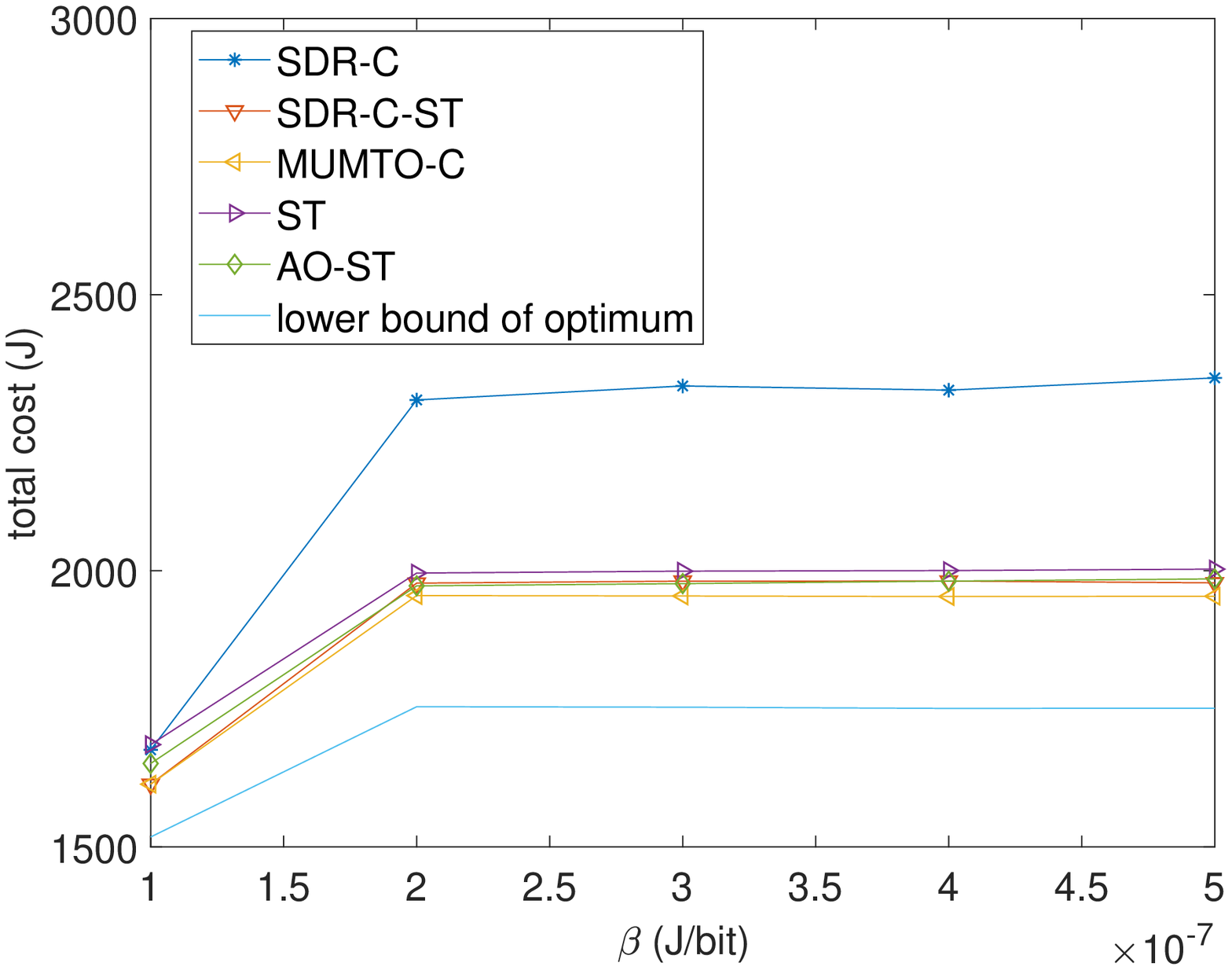}
\caption{Total system cost versus $\beta$ with CAP.}
\label{fig:beta_SDR_CAP}
\includegraphics [scale =0.42]{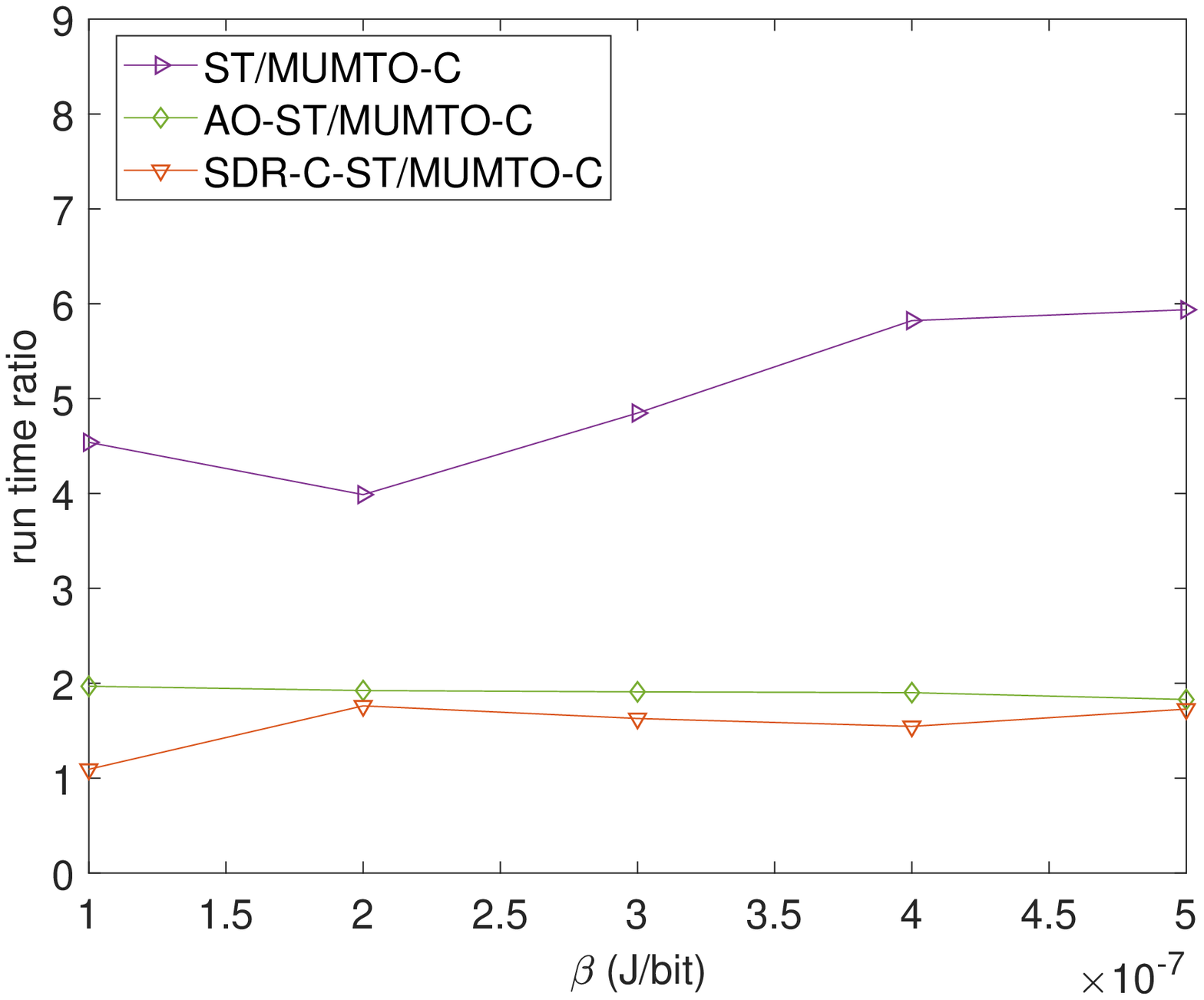}
\caption{Run time ratio versus $\beta$ with CAP.}
\label{fig:beta_time_SDR_CAP}
\end{figure}

To demonstrate the role and contribution of each step in the
MUMTO-C algorithm, we first compare it with the
following methods: 1) the \textit{SDR-C} method where only the
first step of MUMTO-C is applied, 2) the \textit{SDR-C-ST}
method where the AO step is skipped, 3) the \textit{AO-ST} method
where only the last two steps of MUMTO-C are applied by using
random offloading decisions for all tasks as the starting point for
the iterations of AO, 4) the \textit{ST} method where only the last
step of MUMTO-C is applied by using random offloading
decisions for all tasks as the starting point for the iterations of
ST, and 5) the \textit{lower bound of optimum}, which is obtained
from the optimal objective value of the SDR lower bound of problem
\eqref{Eq_new_objective_CAP}.

We show the system cost and the run time ratio vs. $\alpha$ in
Figs.~\ref{fig:alpha_SDR_CAP} and \ref{fig:alpha_time_SDR_CAP},
respectively. Since $\alpha$ is the weight on the CAP usage cost,
more tasks compete at the CAP when $\alpha$ is smaller. We observe
that MUMTO-C can reduce the system cost by up to
20$\%$ compared with purely applying SDR-C and is much closer to
the lower bound of optimum with CAP. Furthermore, though SDR-C-ST,
AO-ST, and ST can provide similarly low cost as
MUMTO-C, which can be attributed to the sequential
searching of ST, they require much longer run time to obtain their
solutions. This demonstrates that we require both the SDR-C and AO
steps in the proposed algorithm to provide an effective starting
point for the ST step to reach a local minimum solution quickly.

\begin{figure}[t!]
\vspace{-0.3cm}
    \centering
\includegraphics [scale =0.42]{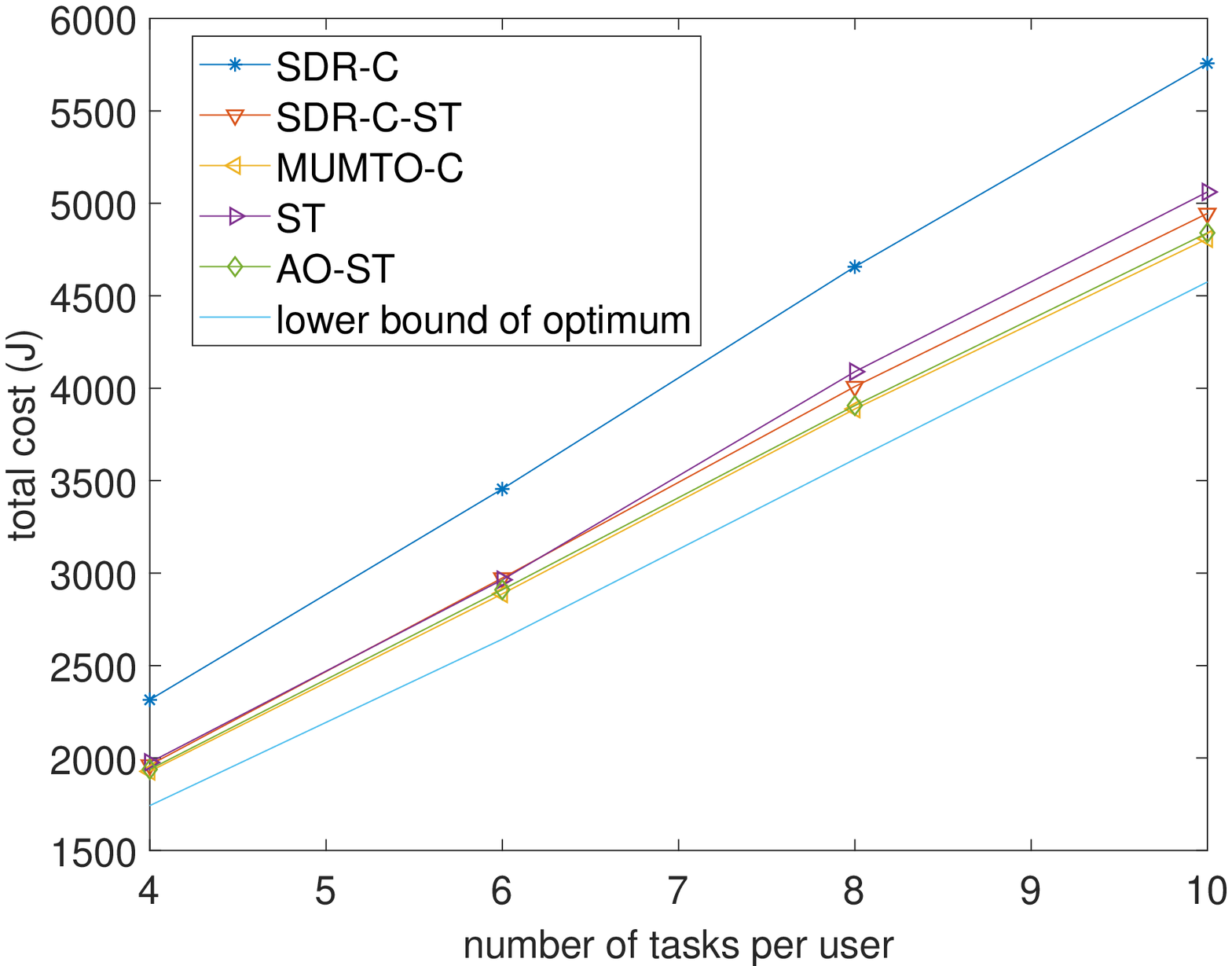}
 \caption{Total system cost versus the number of tasks $M$ per user with CAP.}
\label{fig:task_SDR_CAP}
\includegraphics [scale =0.42]{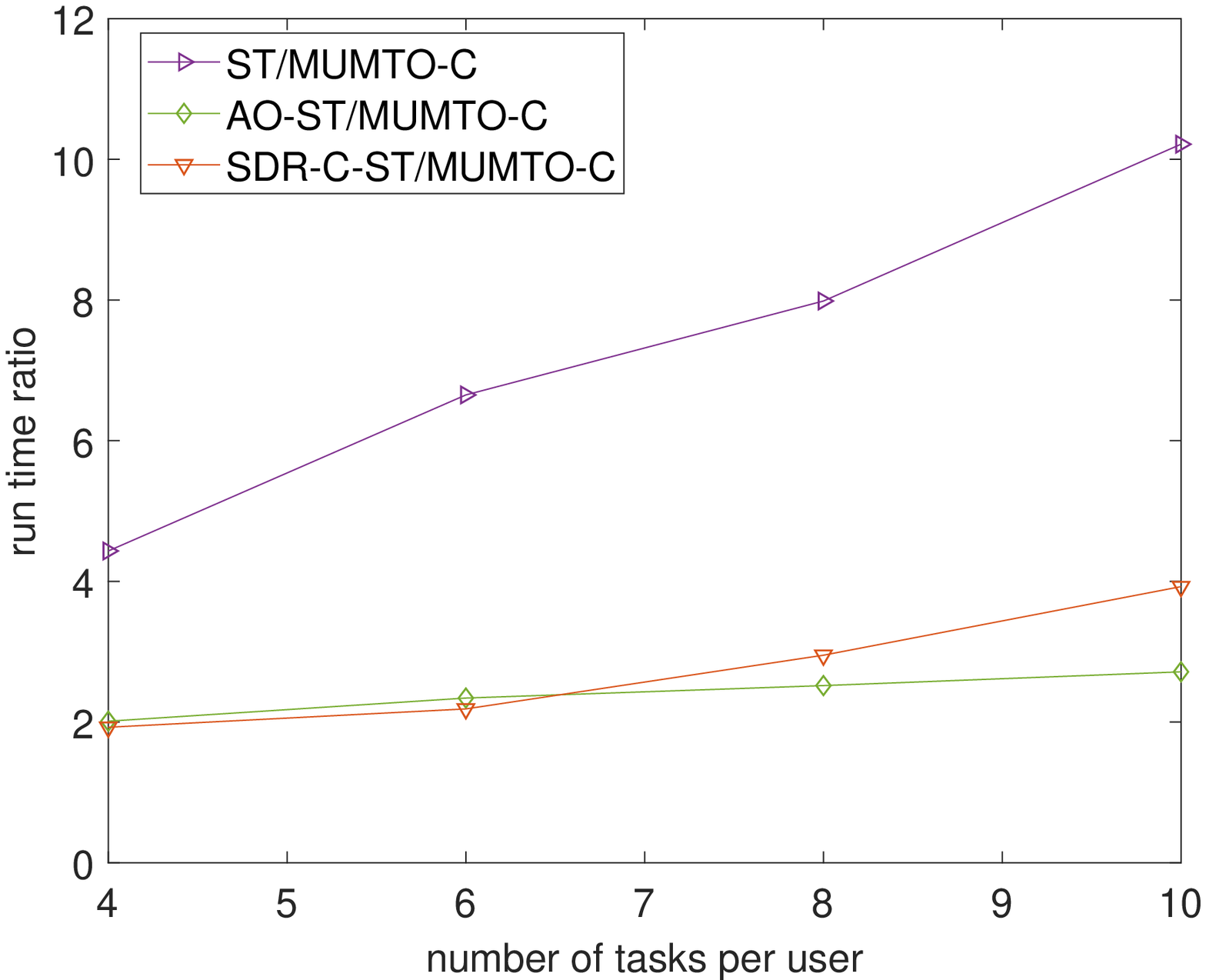}
 \caption{ Run time ratio versus the number of tasks $M$ per user with CAP.}
\label{fig:task_time_SDR_CAP}
    \vspace{-.2cm}
\end{figure}

Similar observations can be made in Figs.~\ref{fig:beta_SDR_CAP} and
\ref{fig:beta_time_SDR_CAP}, where we show the system cost and the
run time ratio vs. the weight $\beta$ on the cloud usage cost, and
in Figs.~\ref{fig:task_SDR_CAP} and \ref{fig:task_time_SDR_CAP},
where we show the system cost and the run time ratio vs. $M$, the
number of tasks per user. When $\beta$ is large, all tasks are more
likely to be processed by either the mobile users or the CAP. More
importantly, MUMTO-C is shown to be more scalable,
since the run-time ratios are nearly linearly increasing with the
number of tasks per user.

\subsubsection{\textbf{Comparison with Further Alternatives}}
\begin{figure}[t!]
\vspace{-0.3cm}
    \centering
\includegraphics [scale =0.42]{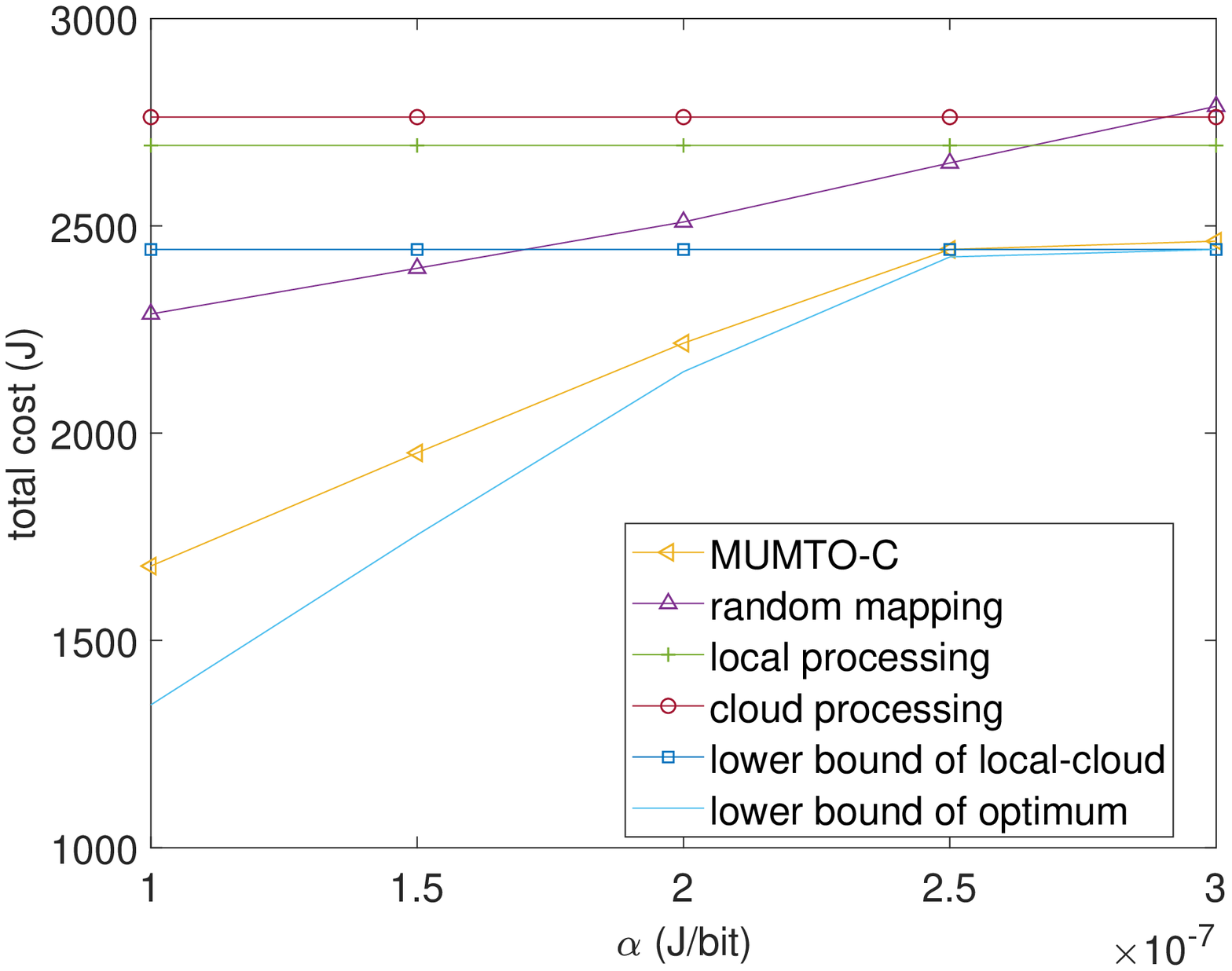}
\caption{ Total system cost versus $\alpha$ with CAP.}
\label{fig:alpha_CAP}
\includegraphics [scale =0.42]{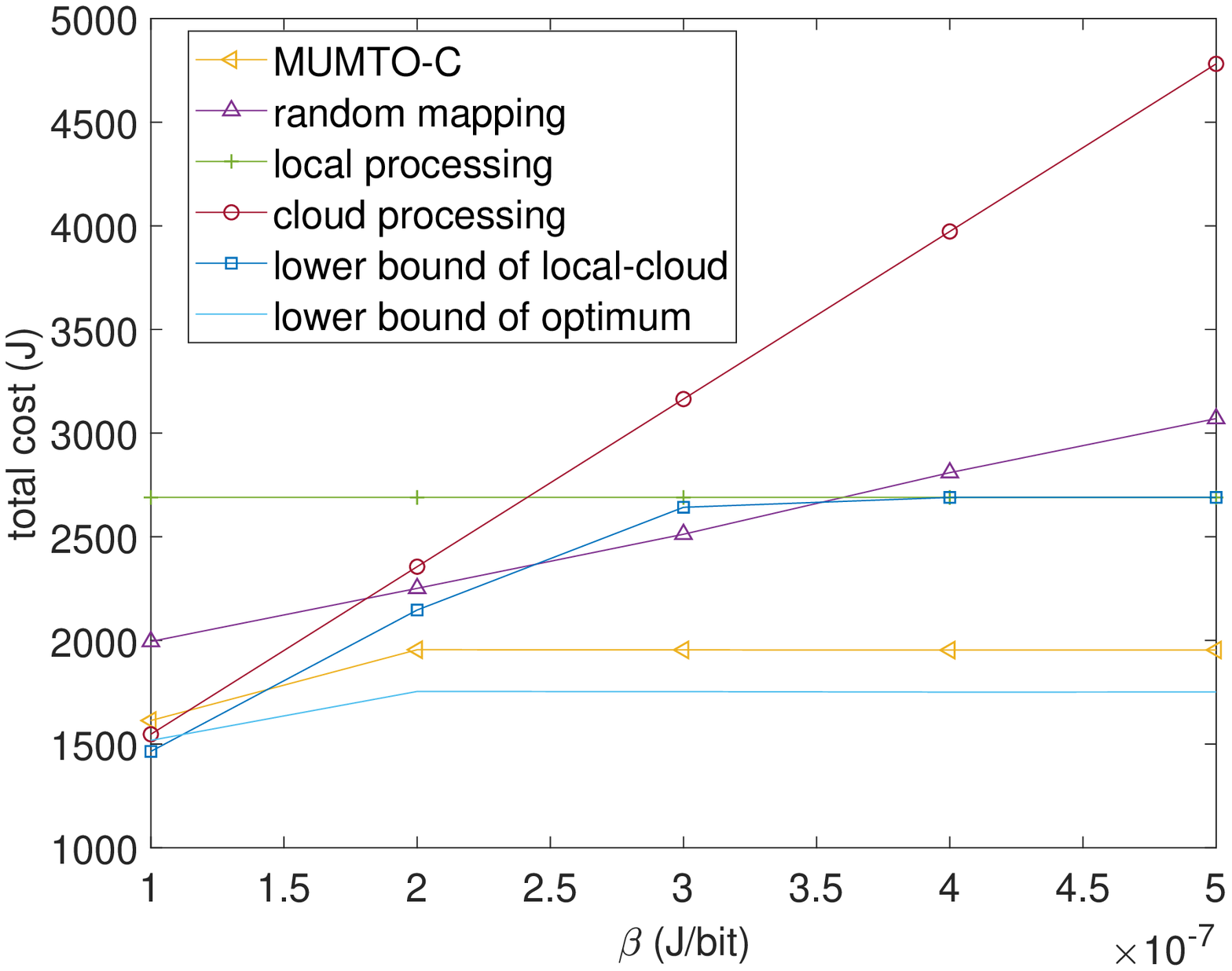}
\caption{Total system cost versus $\beta$ with CAP.}
\label{fig:beta_CAP}
    \vspace{-.2cm}
\end{figure}

 For further performance evaluation, we also consider
the following schemes: 1) the \textit{local processing only} scheme, 2) the
\textit{cloud processing only} scheme, 3)
the \textit{lower bound of local-cloud}, which is the same as \textit{lower bound of optimum}
defined in Sec.~\ref{sec_simulation_noCAP},
and 4) the \textit{random mapping} scheme where each task is
processed at different locations with equal probability. As shown in
Figs.~\ref{fig:alpha_CAP} and \ref{fig:beta_CAP}, the lower bound of
optimum with CAP converges to the lower bound of local-cloud as
$\alpha$ becomes large and the lower bound of local-cloud converges
to the local only method as $\beta$ becomes large. In both figures,
MUMTO-C is near-optimal and substantially outperforms
all alternatives especially when the cost of using the CAP is small
or the cost of using the cloud is large.

\section{Conclusion}
In this work, we have considered a general mobile cloud computing
system consisting of multiple users and one remote cloud server,
where each user has multiple independent tasks. To minimize a
weighted total cost of energy, computation, and the delay of all
users, we aim to find the overall optimal tasks offloading decisions
and communication resource allocation. We show that the resultant
optimization problem is a non-convex separable QCQP. The proposed
MUMTO algorithm uses SDR and binary recovery to jointly compute the
offloading decision and communication resource allocation. For the
scenario with the presence of a CAP, the resultant optimization
problem is even more complicated. We further propose a three-step
MUMTO-C algorithm, which always compute a locally optimal solution.
By comparison with a lower bound of the minimum cost for both
scenarios, we show that both MUMTO and MUMTO-C give nearly optimal
performance and can substantially out perform existing alternatives
over a wide range of parameter settings. Finally, we remark that
there are several interesting directions  for future study, such as
scheduling
 tasks with
strict delay constraints, user mobility  and its impact on the offloading and resource allocation, designing improved methods to better handle dynamically
arriving tasks,  and investigating into  a more general scenario with multiple CAPs.


\end{document}